\documentclass[aps,prx,twocolumn,showpacs,floatfix,nobibnotes,nofootinbib,superscriptaddress,amsmath,amssymb,longbibliography]{revtex4-1}
\usepackage[english]{babel}
\usepackage[colorlinks,urlcolor={blue}]{hyperref}
\usepackage{eurosym}
\usepackage[usenames]{color}
\usepackage{graphicx}
\usepackage{amsmath}
\usepackage{amsfonts}
\usepackage{amssymb}
\usepackage{mathrsfs}
\usepackage{bm}
\usepackage{verbatim}
\usepackage{ulem}

\usepackage[textsize=small]{todonotes}

\newcommand{\beq}{\begin{equation}}
\newcommand{\eeq}{\end{equation}}
\newcommand{\bea}{\begin{eqnarray}}
\newcommand{\eea}{\end{eqnarray}}

\newcommand{\Tlab}{T_{\text{Lab}}}

\newcommand{\nn}{NN}
\newcommand{\nnn}{NNN}

\begin{document}

\title{Power counting in chiral effective field theory and nuclear binding}
\author{C.-J.~Yang}
\email{chiehjen@chalmers.se}
\affiliation{Department of Physics, Chalmers University of Technology, SE-412 96
G\"oteborg, Sweden}
\author{A. Ekstr\"om}
\affiliation{Department of Physics, Chalmers University of Technology, SE-412 96
G\"oteborg, Sweden}
\author{C. Forss\'en}
\affiliation{Department of Physics, Chalmers University of Technology, SE-412 96
G\"oteborg, Sweden}
\author{G. Hagen}
\affiliation{Physics Division, Oak Ridge National Laboratory, Oak Ridge, Tennessee 37831,
USA}
\affiliation{Department of Physics and Astronomy, University of Tennessee, Knoxville,
Tennessee 37996, USA}
\date{\today }

\begin{abstract}
  Chiral effective field theory ($\chi$EFT), as originally proposed by
  Weinberg, promises a theoretical connection between low-energy
  nuclear interactions and quantum chromodynamics (QCD). However, the
  important property of renormalization-group (RG) invariance is not
  fulfilled in current implementations and its consequences for
  predicting atomic nuclei beyond two- and three-nucleon systems has
  remained unknown. In this work we present a first and systematic
  study of recent RG-invariant formulations of $\chi$EFT and their
  predictions for the binding energies and other observables of
  selected nuclear systems with mass-numbers up to $A =
  16$. Specifically, we have carried out \textit{ab initio} no-core
  shell-model and coupled cluster calculations of the ground-state
  energy of $^3$H, $^{3,4}$He, $^{6}$Li, and $^{16}$O using several
  recent power-counting (PC) schemes at leading order (LO) and
  next-to-leading order (NLO), where the subleading interactions are
  treated in perturbation theory. Our calculations indicate that
  RG-invariant and realistic predictions can be obtained for nuclei
  with mass number $A \leq 4$. We find, however, that $^{16}$O is
  either unbound with respect to the four $\alpha$-particle threshold,
  or deformed, or both. Similarly, we find that the $^{6}$Li
  ground-state resides above the $\alpha$-deuteron separation
  threshold. These results are in stark contrast with experimental
  data and point to either necessary fine-tuning of all relevant
  counterterms, or that current state-of-the-art RG-invariant PC
  schemes at LO in $\chi$EFT lack necessary diagrams---such as
  three-nucleon forces---to realistically describe nuclei with mass
  number $A>4$.
\end{abstract}

\pacs{12.39.Fe, 25.30.Bf, 21.45.-v, 21.60.Cs }
\maketitle

\vspace{10mm}

\section{Introduction}
Effective field theory (EFT)~\cite{WEINBERG1979327} provides a
theoretical framework for predicting physical phenomena---normally
within some energy domain of interest---without knowing or assuming
the full details of the underlying physics. Indeed, most physical
systems exhibit many characteristic energy and length scales, and with
the tools of EFT we can exploit such scale separations for analyzing
physical processes. Although it is not always obvious, this approach
is used throughout the physical sciences. For instance, not much is
gained by including the quark degrees of freedom in the quantum electrodynamic
description of the hydrogen atom.

The EFT philosophy appears particularly suitable for application in
low-energy nuclear physics calculations. Indeed, computing nuclei
directly from the Lagrangian of quantum chromodynamics (QCD), via
lattice QCD methods, is extremely complicated, and in most cases
computationally challenging or intractable, in particular within the
non-perturbative region of QCD.

Chiral effective field theory
($\chi$EFT)~\cite{We90,We91,bira,bira1,Epel,Epel1,idaho,idaho1,Epelmore,Epelmore2,reviews,reviews2,reviews3,rev1}
promises a viable method for deriving the low-energy description of
the pion-mediated nuclear interaction that is also constrained by the
symmetries of QCD, and in particular the spontaneous breaking of the
approximate chiral symmetry of quarks. This approach could potentially
connect the description of atomic nuclei to the Standard Model of
particle physics. Furthermore, an EFT offers a handle on estimating
the impact of omitted higher-order dynamics that also contribute to
the epistemic uncertainty of the approach. If the $\chi$EFT
description of the nuclear interactions complies with all
field-theoretical requirements, in particular renormalization group
(RG) invariance, it could significantly increase the predictive power
of \textit{ab initio} computations of nuclear
properties~\cite{dickhoff2004,lee2009,bogner2010,barrett2013,soma2013,hagen2014,HERGERT2016165,
  RevModPhys.87.1067, BARNEA1999427, GLOCKLE1996107}. In this paper we
present a first study of the nuclear binding mechanism in selected
low-mass nuclei using RG-invariant formulations of the strong nuclear
interaction.

The overarching strategy in $\chi$EFT is to start from an effective
Lagrangian including all interaction terms with the same symmetries as
QCD below the chiral symmetry-breaking scale $\sim 1$ GeV. Applying
the methods of chiral perturbation theory yields a potential
description of the inter-nucleon interaction in terms of irreducible
multi-pion exchanges and zero-range contact interactions. In this
sense, $\chi$EFT is often viewed as a low-energy expansion of QCD,
dressed in the relevant degrees of freedom---pions and nucleons---and
sometimes the lowest excitation of the nucleon, i.e. the
$\Delta(1232)$-isobar~\cite{bira1,Hemmert_1998,Kaiser:1998wa,Krebs:2007rh}.

In EFT studies of nuclei, one aims at predicting low-energy nuclear
observables using an order-by-order improvable potential-expansion in
terms of a small parameter constructed as a ratio between the
physically relevant soft and hard scales. In $\chi$EFT, the hard
momentum-scale is $\Lambda_b \approx 0.5-1$ GeV, and the soft scale is
$Q={\rm max}(q,m_{\pi})$, where $q$ denotes the external
(initial/final state) momentum scale of the interacting nucleons, and
$m_{\pi} \sim 140$ MeV denotes the pion mass. Up to a certain order
$\nu$ in the chiral expansion, only a finite number of interaction
terms, or diagrams, contribute. The organizational scheme for
assigning a diagram to a specific order in the EFT according to its
expected importance is referred to as \textit{power counting} (PC). Besides
renormalizing the potential, the PC should also ensure RG-invariant
amplitudes, i.e. observables. RG-invariance is a crucial requirement
in any EFT. When integrating out the pion degree of freedom, one can
construct a so-called pionless EFT, which is easier to deal with
analytically and for which the path towards renormalizability is
clear~\cite{VANKOLCK1999273}. This framework has been employed for
successfully describing few-nucleon systems, predominantly helium
isotopes~\cite{PLATTER2005254,Kirscher:2010vg,PhysRevC.92.054002,PhysRevC.94.034003},
and for extrapolating lattice QCD
predictions~\cite{PhysRevLett.114.052501}. However, pionless EFT
appears inadequate for predicting realistic properties of light- and
medium-mass nuclei heavier than
$^{4}$He~\cite{STETCU2007358,CONTESSI2017839,PhysRevC.98.054301,pionlessfail},
and it remains an open question whether subleading orders will provide
a remedy.

In nuclei, the likely importance of inter-nucleon interactions with
external momenta $q \gtrsim m_{\pi}$ suggests the need for an explicit
inclusion of pion physics and the use of $\chi$EFT. Unfortunately, the
presence of the pion propagator typically complicates the
Schr\"odinger equation to the extent that analytical studies become
intractable. One must therefore resort to numerical checks of
RG-invariance at each chiral order. For nuclear structure
calculations---which are always performed within a truncated Hilbert
space---enlarging the model space will determine whether all
high-momentum (short range) dynamics are properly accounted for as
contact interactions. In practice, this is typically done by
increasing the imposed momentum-cutoff ($\Lambda$) that serves to
regularize the potential. In this procedure, additional high-momentum
details are explicitly exposed. RG-invariance is destroyed if the
short-range couplings (counterterms)---typically referred to as
low-energy constants (LECs)---fail to run with the additional
high-momentum ingredients.  The resulting lack of RG-invariance yields
observable predictions that depend on the regularization procedure. In
contrast, an EFT is order-by-order renormalizable if the predicted
observables evaluated up to order $\nu$ have residual cutoff
dependence equal or less than $(\frac{Q}{\Lambda})^{\nu+1}$.


Chiral perturbation theory provides an order-by-order renormalizable
framework for constructing a low-energy EFT of QCD, and has also been
applied quite successfully to the single-nucleon sector with explicit
pions, see e.g. Refs~\cite{chpt1,chpt2}.
Problems emerge, however, with the inclusion of two or more
nucleons. These difficulties were not entirely clear in the early days of
$\chi$EFT~\cite{We90} as it was initially assumed that the PC employed
in the single-nucleon sector would successfully carry over to
renormalize also the multi-nucleon sector. This approach is
colloquially referred to as Weinberg Power Counting (WPC) and is the
\textit{de facto} PC employed in quantitatively realistic descriptions
of atomic nuclei. Nevertheless, it is already well known that $\chi$EFT
based on WPC will not generate RG-invariant results for observables, see
e.g. Ref.~\cite{nogga}.

Nowadays, there exist several PCs for
$\chi$EFT~\cite{BY3p0,BYtri,BYs,1s0d,Birse,Birse1,Birse2,Valdper,Valdper1,BY,vald17,bingwei18}
that produce RG-invariant nucleon-nucleon (\nn{}) scattering
amplitudes. However, in the present paradigm of \textit{ab initio}
computations, such PCs remain unexplored in studies of atomic nuclei
with mass number $A \geq 4$. In fact, there are merely two attempts to
demonstrate RG invariance of nuclear structure calculations beyond the
\nn{} sector. These are Faddeev-type calculations of the three-body
systems $^3$H,$^3$He~\cite{nogga,song}.

In this work we significantly broaden the established field of
low-energy nuclear theory by applying RG-invariant $\chi$EFT
interactions to selected nuclei with mass numbers $A\leq 16$. This
constitutes an important leap forward in the exploration of
RG-invariant formulations of $\chi$EFT~\cite{BYs}.
We consider some of the most recent RG-invariant $\chi$EFT
formulations~\cite{BY3p0,BYtri,BYs}, and employ the no-core shell
model (NCSM)~\cite{ncsm,ncsm1} and the coupled-cluster (CC)
method~\cite{cc,cc1,cc2,cc3,cc4,cc5,hagen2014} to calculate the
ground-state energy and nuclear charge radius of $^{3}$H, $^{3}$He,
$^{4}$He, $^{6}$Li, and the ground-state energy of $^{16}$O,
respectively. Note that the total binding energy for the system is the
negative of the ground-state energy.
The NCSM gives, in principle, an exact solution to the many-nucleon
Schr\"odinger equation, but is limited to light nuclei due to the
exponential increase in computing cost with the system size (a
combined measure of the number of basis states and nucleons). On the
other hand the CC method has a much softer (polynomial) scaling with
the system size, and it gives a controlled and systematically
improvable approximation to the exact solution for the
wavefunction~\cite{cc5,hagen2014}. The access to consistently
increasing computational power and the development of similarity RG
techniques~\cite{bogner2010} enables computation of nuclei as heavy as
$^{100}$Sn using \textit{ab initio} CC and similar
methods~\cite{simonis2017,PhysRevLett.120.152503,gysbers2019,soma2020}.
In this work, however, we are focusing on RG invariance and must therefore
explore relatively large values for the regulator cutoff in the
interactions. In both NCSM and CC, the size of the employed basis must
go hand-in-hand with a large cutoff to resolve the short-range part of
the interaction while also capturing the long-range part of the
wavefunction. It is therefore a big computational challenge to predict
nuclear many-body observables using a nuclear interaction with strong
high-momentum details. Where possible, we employ recent extrapolation
techniques~\cite{ir,ir0,ir2,ir3,ir4} to obtain reasonably converged
results for the ground-state energy of the nuclei considered in this
work.
 
In a renormalizable $\chi$EFT, the subleading contributions,
i.e. beyond leading order (LO), are treated in perturbation
theory. Indeed, in order to achieve RG-invariance at the \nn{} level
it has been shown that---due to a Wigner bound-like
effect~\cite{wigner,wigner2}---one has to either treat all subleading
contributions perturbatively or promote at least two short-range contact terms
non-perturbatively at the same
time~\cite{oller,oller2,oller3}\footnote{There also exists a
  renormalization scheme which corresponds to using infinitely many
  contact terms~\cite{ge2,de8}.}. In this work we will follow the
strategy of perturbatively including subleading contributions, and we
also demonstrate how a Hellmann-Feynman procedure can be used to
achieve this without modifying existing many-body solvers.

This manuscript is organized as follows: In Sec.~\ref{sec:MWPC}, we
briefly introduce the $\chi$EFTs used in this work and Weinbergs'
initial approach. Then, in Sec.~\ref{sec:fewbodypred}, we present
\textit{ab initio} predictions for $^3$H and $^{3,4}$He up to NLO in a
well-known RG-invariant PC. Corresponding NCSM and CC calculations for
$^{6}$Li and $^{16} $O, respectively, are presented in
Sec.~\ref{sec:manybodypred}. In Sec.~\ref{sec:otherpcs} we describe
some additional and relevant PC schemes based on a dibaryon-field, a
separable version of the dibaryon field~\cite{1s0d,sep}, and a
perturbative treatment of most $P$-waves~\cite{bingwei18}.
In this Section we also present the CC predictions for the
ground-state energy in $^{16}$O using such alternative PC schemes.  We
summarize our findings and their implications in
Sec.~\ref{sec:summary}.

\section{Modified Weinberg power counting (MWPC) \label{sec:MWPC}}
Detailed properties of several nuclear systems can nowadays be
successfully described by solving the non-relativistic Schr\"{o}dinger
equation using sophisticated potentials based on Weinberg's initial
approach~\cite{We90,We91}, see
e.g. Refs.~\cite{PhysRevLett.120.152503,Hagen:2015yea,Elhatisari:2015ug,PhysRevLett.122.042501,abn3lo,abn3lo1,abn3lo3,abn3lo6,nnloopt,nnlosat,nnlodelta,weiguang20}.
Such interactions also enable a description of low-energy \nn{}
scattering data with an accuracy comparable to~\cite{idaho1} and
beyond~\cite{n4lo,n5lo,Epeln5lo} existing high-precision and
phenomenological potentials~\cite{CdBonn,av18,nij}. Despite the fact
that WPC has enabled successful \textit{ab initio} models of the
strong nuclear interaction, there are several reasons for modifying
Weinberg's initial prescription~\cite{We90,We91} for generating \nn{}
and three-nucleon (\nnn{}) potentials. In particular, the amplitude produced
from Weinberg's prescription for generating the potential is not RG
invariant~\cite{nogga,Ya09A,Ya09B,ZE12}. We emphasize that WPC is
important for guiding experimental and theoretical analyses of nuclei
and nuclear systems, but it does not necessarily lead us closer to
analyzing nuclei from first principles, i.e from QCD. We should also
point out that there exist arguments~\cite{ge,ge2,de3,de4,de8} for an
alternative view on renormalization in $\chi$EFT. See e.g.
Refs.~\cite{Epelbaum:2013tc,bingwei_rev,de5,de6,de7,yangreview,yang20,rev1,rev2}
for extensive discussions on opposing views regarding this
topic. Besides the above problems, Weinberg’s prescription also lacks a pion-mass-dependent contact term---which is demanded by RG in the chiral extrapolation applications~\cite{ksw,ksw1}. 
Although this is an important aspect of the
theory, it is not the focus of our present work.

The conventional implementation of $\chi$EFT proceeds in two
steps. First, one constructs the long-range (pion-exchange) potential
from the chiral Lagrangian. Then one collects the necessary
short-range diagrams into a contact potential to cancel the
divergences of the aforementioned long-range pion potential, and
subsequently iterates the sum of all potential terms non-perturbatively
in the Lippmann-Schwinger or Schr\"odinger equation to obtain the
amplitudes for constructing e.g. the scattering $S$-matrix. The
resulting potentials are singular at short distances (large momenta),
and therefore require regularization using a regulator function
$f_{R}$ with an ultraviolet cutoff $\Lambda$. In this work we use a
momentum-space representation and employ a standard, non-local,
regulator function
\begin{equation}
  f_R(p;\Lambda )=\exp \left[ \left(-p / \Lambda \right)^{2n} \right],
\end{equation}
with $n=2$.
We denote the initial (final) relative momenta with
$\mathbf{p}$ $(\mathbf{p}')$, and use $\mathbf{q}=\mathbf{p}'-\mathbf{p}$
for the momentum transfer. Note that local regulators can also be
adopted, and this has been explored in coordinate-space quantum Monte
Carlo calculations up to NNLO in $\chi$EFT using
WPC~\cite{Gezerlis:2013ipa,Gezerlis:2014zia}. One can also mix the
local and non-local formalism~\cite{Epelmore2,deannew}. See also
Ref.~\cite{Dyhdalo:2016ygz} for a detailed discussion of some of the
observed artifacts induced by different regulator functions.

Clearly, predictions of observables should not depend on the chosen
regulator or the value of the regulator cutoff $\Lambda$, i.e. the
LECs in the contact potential must act as counterterms and run with
$\Lambda$ at each chiral order. To achieve RG-invariant amplitudes we
must modify WPC.

\subsection{Leading order}
At LO in WPC, the interaction potential consists of the well-known
one-pion-exchange potential (OPE) accompanied by two \nn{} contact
terms acting in the singlet and triplet $S-$waves. In
momentum-space it is represented as
 \begin{equation}
   V_{\rm LO}^{\rm WPC}(\mathbf{p},\mathbf{p}')=
   \frac{g_A^2}{4f_{\pi}^2}\mathbf{\tau}_1\cdot\mathbf{\tau}_2\frac{(\mathbf{\sigma}_1\cdot
     \mathbf{q})(\mathbf{\sigma}_2\cdot \mathbf{q})}{m_{\pi}^2 +
     \mathbf{q}^2} + \tilde{C}_{^1S_0} + \tilde{C}_{^3S_1}.
 \end{equation}
Here, $\tilde{C}_{^1S_0},\tilde{C}_{^3S_1}$ denote the LO contact LECs
acting in separate partial waves. Above, and in the following, we
suppress the $\Lambda$-dependence of the LECs. Also, we adopt the
value $g_{A}=1.27$ for the axial coupling and $f_{\pi }=93$ MeV for
the pion decay constant, respectively, and employ $m_{\pi }=138$ MeV
and $m_{N}=938.9$ MeV for the pion and nucleon mass, respectively.

\begin{figure}[t]
\includegraphics[width=\columnwidth]{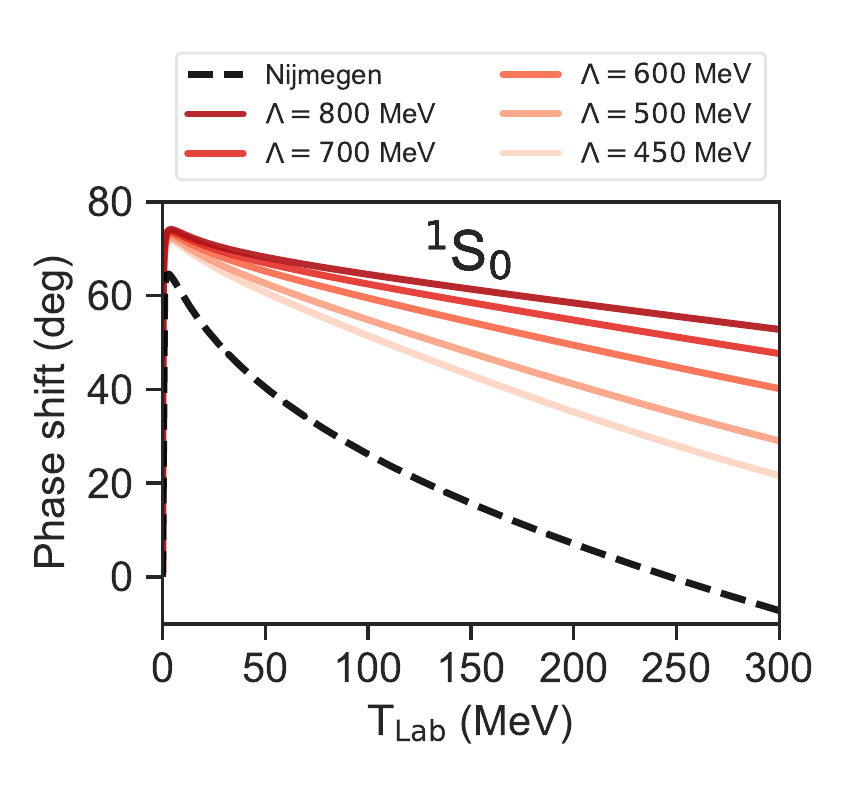}
\caption{The $^{1}S_{0}$ \nn{} scattering phase shifts at LO in MWPC
  as a function of laboratory scattering energy $\Tlab$ for several
  values of the regulator cutoff $\Lambda$. Note that the
  over-attraction persists even for the lowest cutoff value
  $\Lambda=450$ MeV.}
\label{fig:ph1s0}
\end{figure}

It is well-established that $V_{\rm LO}^{\rm WPC}$ produces
non-renormalizable amplitudes in the singular and attractive
partial-wave channels , e.g., $^3P_0$ and
$^{3}P_{2}-^{3}F_{2}$~\cite{nogga}. A remedy of this situation can be achieved by promoting two
additional contact terms at the potential level (otherwise subleading
in WPC) to the $^3P_0$ and $^{3}P_{2}$ channels (one for each), and by
treating all partial waves with angular-momentum quantum number
$\ell>1$\footnote{In Sec.~\ref{sec:perp} we explore a PC with a
  perturbative treatment of most $P$-waves.}
perturbatively~\cite{bingwei18}. These modifications will lead to
RG-invariant \nn{} amplitudes at LO . Due to its similarity with WPC at LO,
we will refer to this RG-invariant PC as modified Weinberg power
counting (MWPC) throughout this work. The corresponding momentum-space
potential at LO is given by
\begin{equation}
  V_{\rm LO}^{\rm MWPC}(\mathbf{p},\mathbf{p}') = V_{\rm LO}^{\rm
    WPC}(\mathbf{p},\mathbf{p}') + (\tilde{C}_{^3P_0} +
  \tilde{C}_{^3P_2})pp'.\label{lomwpc}
\end{equation}
At this order we obtain the amplitudes non-perturbatively in all
partial-waves with $\ell \leq 1$. Note that MWPC and WPC coincide with
each other in $S$-waves. Furthermore, the $S-$wave component of the
nuclear interaction has a large impact on nuclear binding energies,
and the \nn{} scattering phase shifts from MWPC in the $^{1}S_0$
partial wave show a sizeable over-attraction with respect to the
Nijmegen partial-wave analysis~\cite{nij} even at very low scattering
energies, see Fig.~\ref{fig:ph1s0}. To remedy this unphysical
over-attraction, it is motivated to consider alternative PCs, which
will be discussed further in Sec.~\ref{sec:otherpcs}. Throughout this
work, we neglect any isospin-breaking contributions in the PC schemes
we employ. However, we do include the Coulomb interaction
$\sim\alpha/r$ ($\alpha$ is the fine structure constant) non-perturbatively at LO in all
\textit{ab initio} calculations although it, in principle, requires
special treatment due to a renormalization issue. See, e.g., the discussion in Ref.~\cite{cou_issue}.

For quantitative predictions we must infer numerical values of the
relevant LECs for every value of the regulator cutoff $\Lambda$ we
employ. The primary goal of this work is to present the first
predictions of bulk properties, primarily the ground-state energy, of
selected atomic nuclei up to $^{16}$O using MWPC, as well as some
other RG-invariant \nn{} interactions. Thus, to proceed with a first
analysis we straightforwardly determine the numerical values for the
contact LECs such that the chosen theory reproduces the experimental
values for a selected set of calibration observables. A future
procedure could entail a more detailed statistical inference analysis
of the underlying EFT uncertainty as well as the LECs
themselves~\cite{PhysRevC.92.024005,bay4,Wesolowski_2016}.

\begin{figure}[t]
\includegraphics[width=1.0\columnwidth]{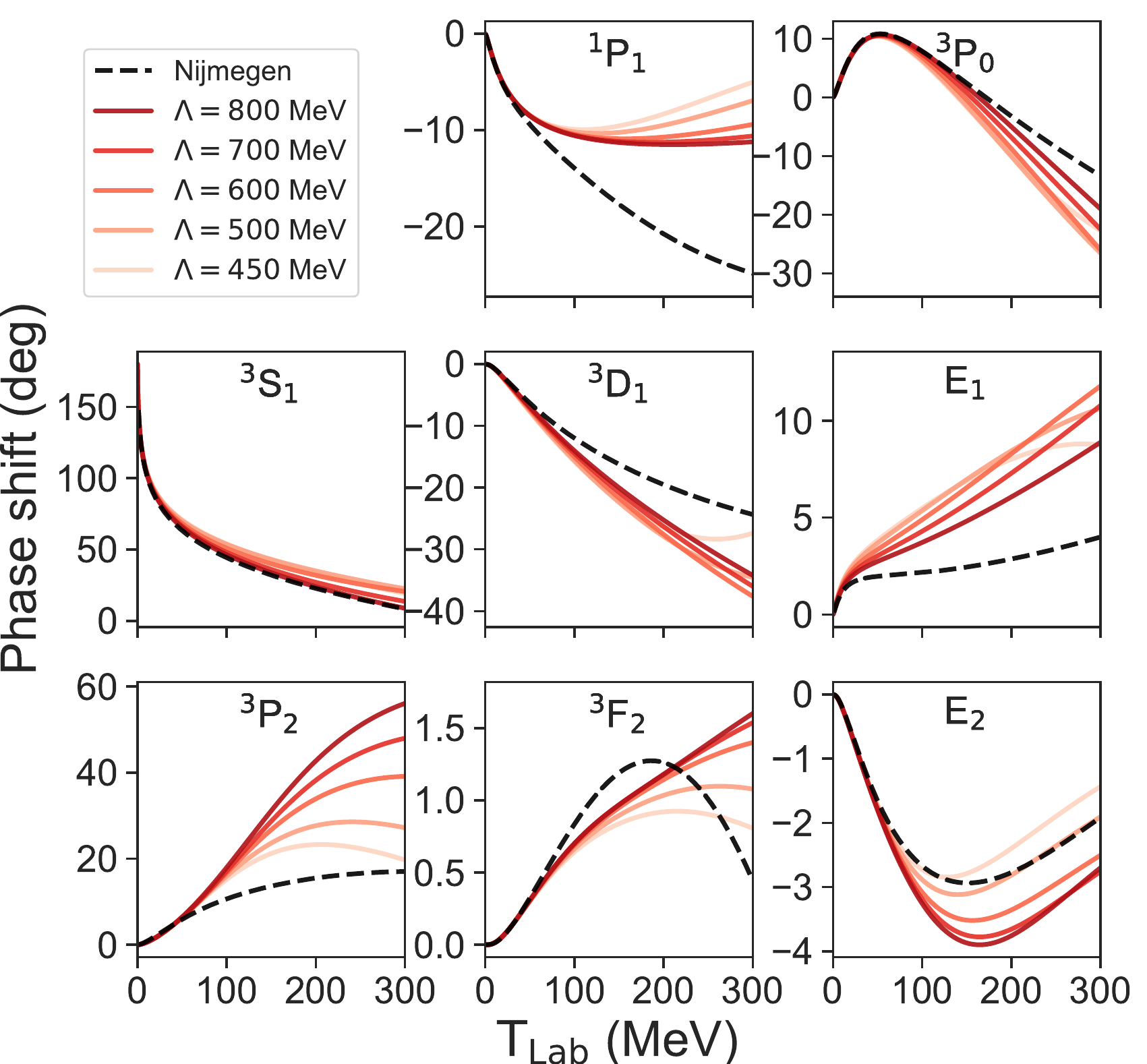}
\caption{Selected phase shifts at LO in MWPC as a function of
  laboratory scattering energy $\Tlab$. Here, the $C_{^3S_1}$ LEC is
  fitted to reproduce the deuteron binding energy while the $P-$wave
  LECs are
  fitted to reproduce the phase shifts at $\Tlab=40$ MeV.}
\label{fig:ph_panel}
\end{figure}

\begin{figure}[t]
\includegraphics[width=\columnwidth]{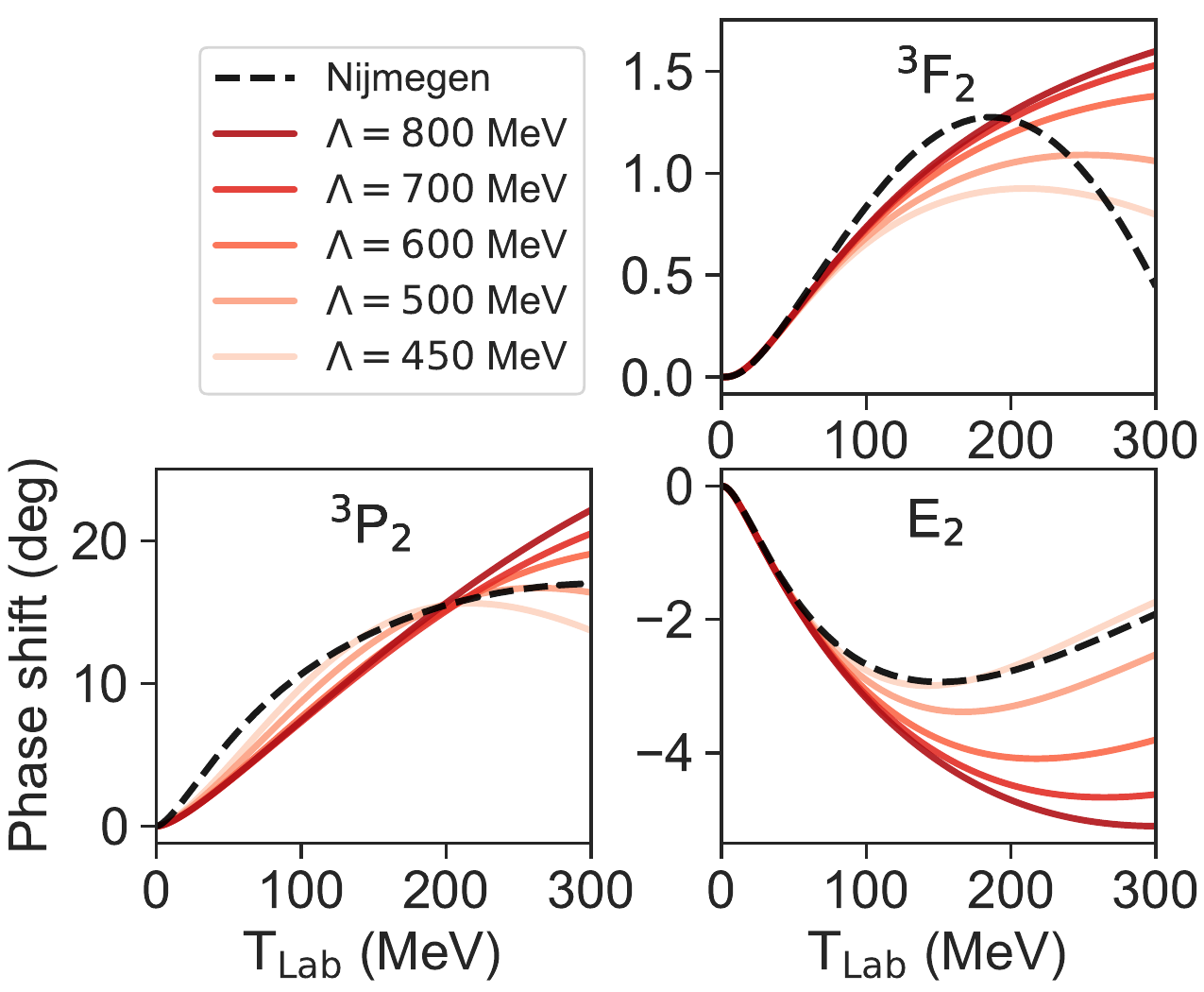}
\caption{$^{3}P_{2}-^{3}F_{2}$ phase shift at LO in MWPC as a function
  of laboratory kinetic energy $\Tlab$. Here the relevant LEC is
  fitted to reproduce the phase shifts at $\Tlab=200$ MeV.}
\label{fig:3pf_200}
\end{figure}

Since we employ a pionful theory, we generally prefer to renormalize
the LECs at a relative \nn{} momentum $k$ corresponding to $m_{\pi }$
where possible. This relative momentum corresponds approximately to a
laboratory scattering energy $\Tlab=$40 MeV. However, to accommodate
the nearly-bound character of the $^{1}S_{0}$ channel we had to pick a
different kinematical calibration point for this channel. Indeed,
matching the only counterterm in this channel to reproduce the phase
shift at $k\sim m_{\pi }$ leads to a rather poor reproduction of the
phase shift at $k<m_{\pi}$. We therefore fit the LO LEC in the
$^{1}S_{0}$ channel to reproduce the $S-$wave scattering length
$a_{0}=-23.7$ fm~\cite{manolo05}. Also, in the $^{3}S_{1}-^{3}D_{1}$
channel, we renormalize the $\tilde{C}_{^3S_1}$ counterterm to reproduce the
deuteron binding energy. For $^{3}P_{2}-^{3}F_{2}$, when calibrating
to reproduce the phase shift at $k\sim m_{\pi}$, we observed a sizable
over-attraction with respect to the Nijmegen analysis for $k>m_{\pi
}$. This is clearly visible in Fig.~\ref{fig:ph_panel}.
To study the impact of this over-attraction, we alternatively fit the
$^{3}P_{2}-^{3}F_{2}$ phase shifts at $\Tlab=200$ MeV. We will refer
to these different LO interactions as MWPC(40) and MWPC(200). The
latter fit yields a dramatically different result for this coupled
channel, see Fig.~\ref{fig:3pf_200}. As expected, and as we will see
in Sec.~\ref{sec:fewbodypred}, this has a negligible, percent-level
impact on the binding energy in few-nucleon systems. On the other
hand, the details of the fitting strategy appears to have a
significant effect on the \textit{ab initio} description of the
ground-state energy in$^{16}$O. This is a key finding of this work,
and we will return to this point in more detail in
Sec.~\ref{sec:manybodypred}.
In MWPC, at cutoff $\Lambda =750$ ($1050$) MeV, deep spurious bound
states start to appear in the $^{3}P_{0}$ ($^{3}S_{1}-^{3}D_{1}$)
channel. We follow the standard projection method as listed in Appendix
B of Ref.~\cite{nogga} to remove those states. Ideally, an
associated parameter $\lambda$, that is used to control the projection
of the spurious states, should be taken very large. However, this
would also result in extremely large values for the matrix
elements in the \textit{ab initio} calculations and will induce
numerical problems. We find it sufficient to
employ $\lambda \approx 10-15$ GeV.
\subsection{Next-to-leading order}
\begin{figure}[t]
\includegraphics[width=\columnwidth]{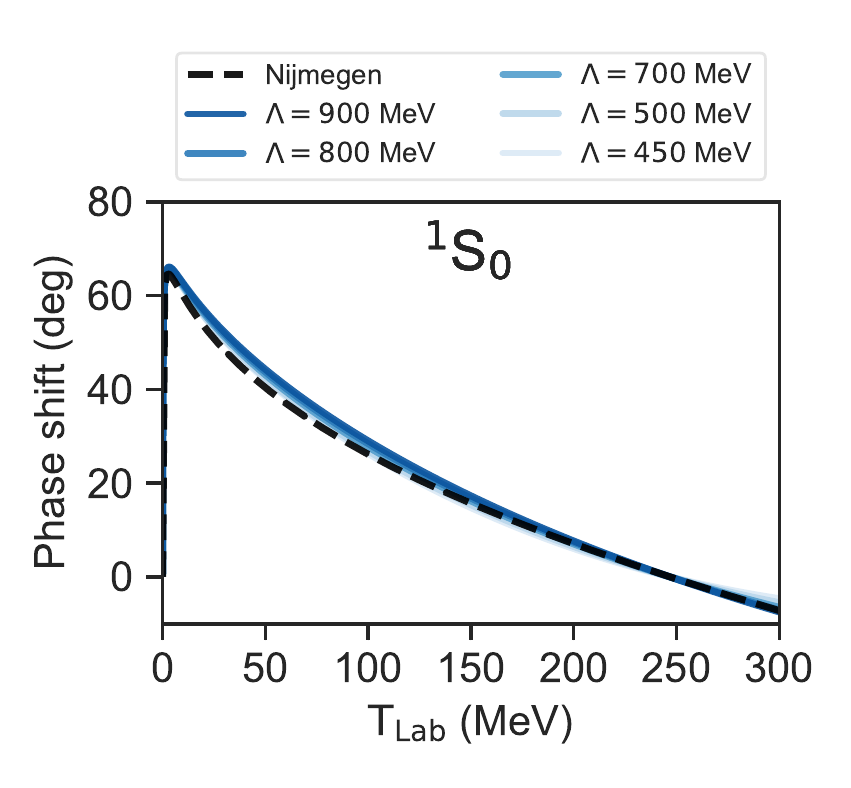}
\caption{The $^{1}S_{0}$ phase shift up to NLO in MWPC as a function
  of laboratory scattering energy $\Tlab$. The relevant LECs are
  fitted to reproduce the scattering length $a_0=-23.7$ fm and
  and the Nijmegen phase shift at $\Tlab=250$ MeV.}
\label{fig:ph_1s0nlo}
\end{figure}

\begin{figure}[t]
\includegraphics[width=\columnwidth]{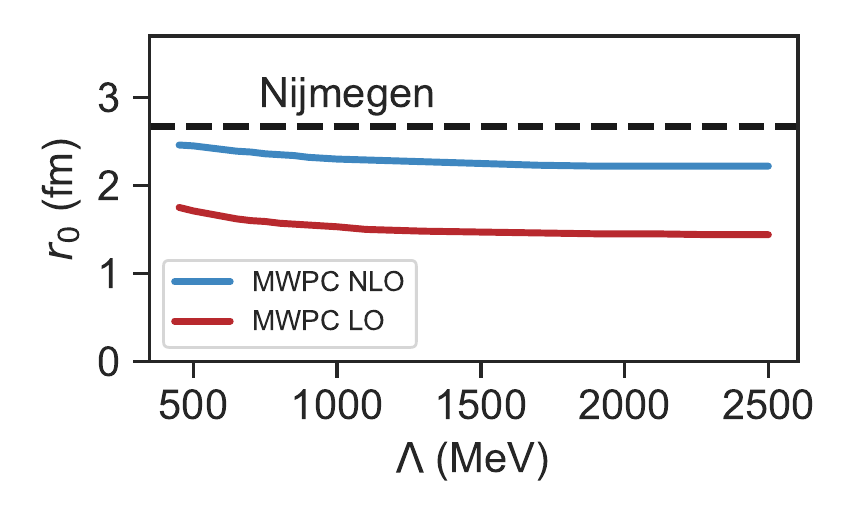}
\caption{Predictions for the effective range $r_0$ in the $^{1}S_{0}$
  channel up to NLO in the MWPC. The MWPC NLO result is closest to the
  Nijmegen value (dashed line).}
\label{plot_wpc_r}
\end{figure}

According to the analyses in
Refs.~\cite{BY3p0,BYtri,BYs,1s0d,BY}, the NLO contribution in MWPC has
to come one chiral order before the appearance of the
two-pion-exchange potential. In fact, the entire NLO contribution to
the amplitude only consists of $^{1}S_{0}$ short-range interactions
\begin{equation}
V_{\rm NLO}^{\rm MWPC}(\mathbf{p},\mathbf{p}')= C_{^1S_0}+\hat{C}_{^1S_0}(p^2+p'^2) \label{nlomwpc}.
\end{equation}
We treat sub-leading orders perturbatively and this $^{1}S_{0}$
contribution is evaluated in the distorted wave Born approximation. To
be clear, there are three differences between the NLO in MWPC and WPC:
\begin{enumerate}
  \item[(i)] NLO in MWPC appears one chiral order earlier than in WPC
  \item[(ii)] NLO in MWPC contains only short-range terms as listed in
    Eq.~(\ref{nlomwpc}).
  \item[(iii)] The NLO interaction is treated perturbatively in MWPC,
    rather than being iterated to all orders as in WPC.
  \end{enumerate}
  We note that the NLO contribution in
Eq.~(\ref{nlomwpc}) contains two additional LECs at NLO, acting only
in the $^{1}S_{0}$ channel. However, the LEC $C_{^{1}S_0}$ is the NLO
correction to $\tilde{C}_{^{1}S_{0}}$. Thus, we effectively only have
two LECs in the $^{1}S_{0}$-channel up to, and including, NLO. Still,
this gives us plenty of freedom to describe the corresponding phase
shift. In this work we choose to renormalize the $^{1}S_{0}$ LEC to
reproduce the scattering length $a_0=-23.7$ fm and the Nijmegen
phase shift at $\Tlab=250$ MeV. As expected, the resulting
predictions exhibits a very nice agreement with the Nijmegen analysis,
as shown in Fig. \ref{fig:ph_1s0nlo}. Note also the very weak
dependence on the regulator cutoff $\Lambda$ at this order
(\textit{cf.} Fig.~\ref{fig:ph1s0}).
The LO and NLO predictions for the effective range $r_0$, as a
function of the regulator-cutoff $\Lambda$, are shown in
Fig. \ref{plot_wpc_r}. Here, we employ a larger range of momentum
cutoffs just to demonstrate the expected plateau-behavior of an
RG-invariant amplitude. In \nn{} calculations it is typically not
challenging to take $\Lambda$ to even larger values, e.g. $10-20$
GeV. However, most \textit{ab initio} methods for solving the
many-body Schr\"odinger equation fail to converge for
$\Lambda \gtrsim 600$ MeV due to strong induced wavefunction
correlations, and limitations on the employed model-space
sizes. Several examples of this will be encountered below.
\section{Predictions for $^{3}$H and $^{3,4}$He using $\chi$EFT potentials in MWPC\label{sec:fewbodypred}}
\begin{figure*}[ht]
  \includegraphics[width=\columnwidth]{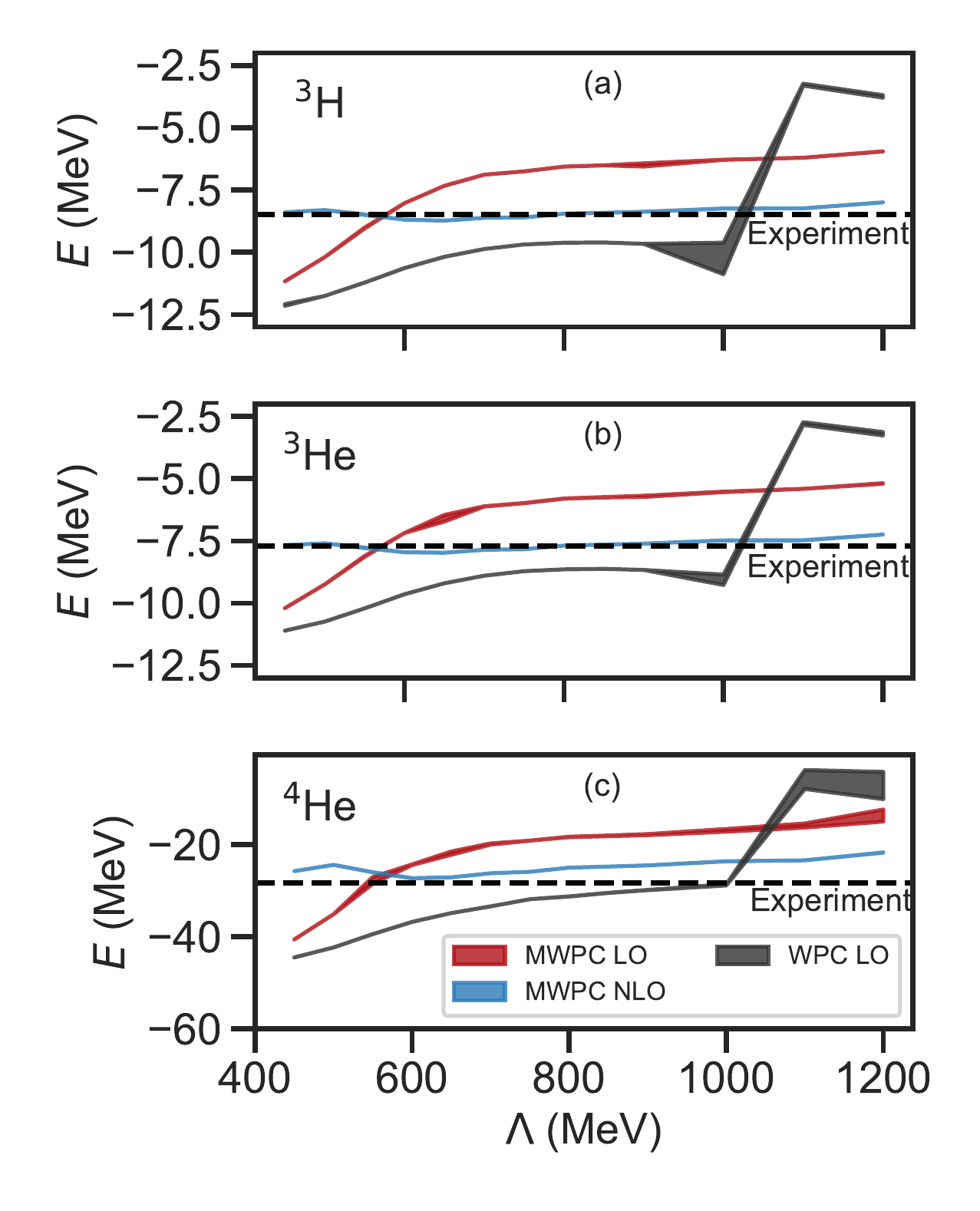}
  \includegraphics[width=\columnwidth]{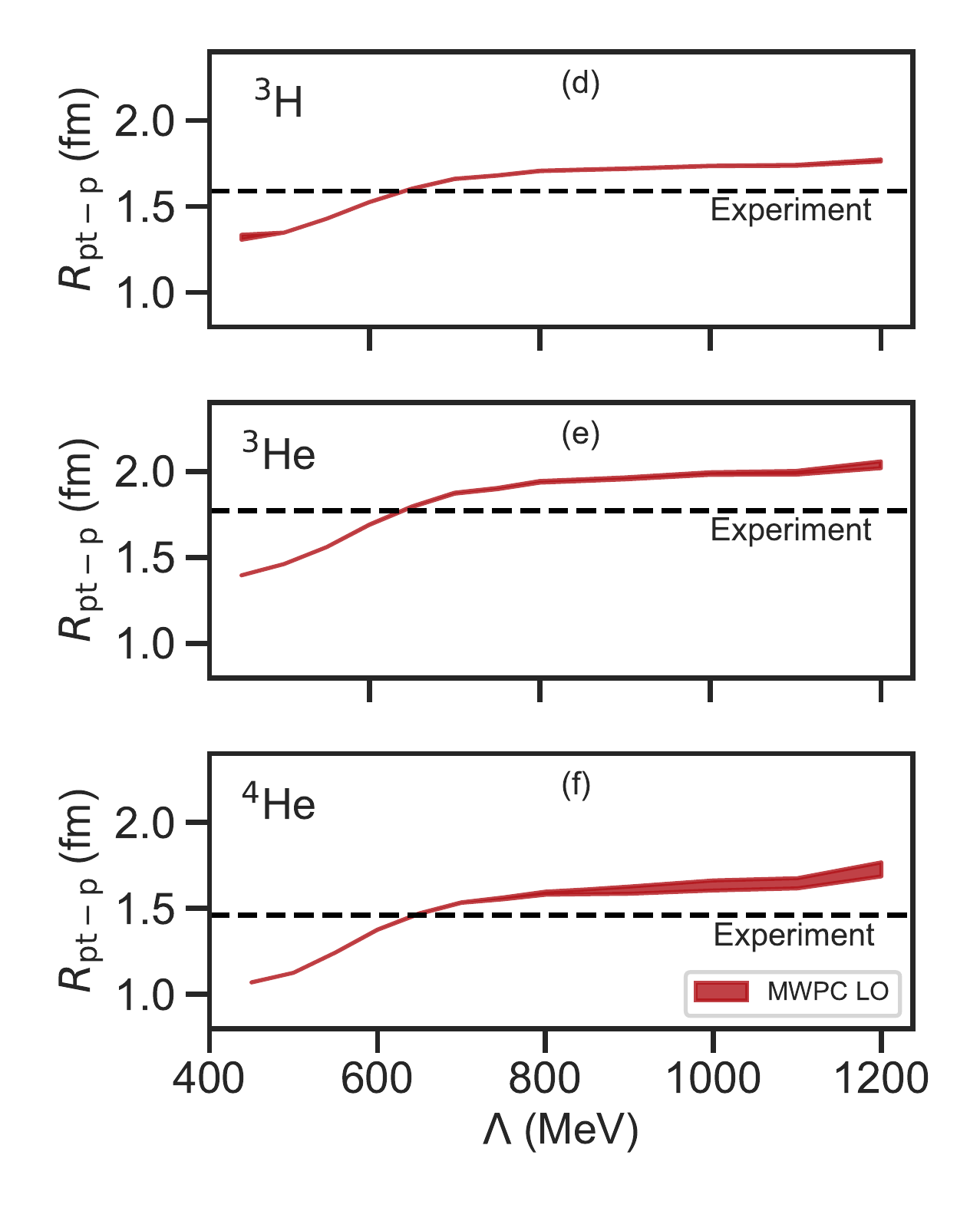}
  \caption{Ground-state energies (panels (a)-(c)) and point-proton
    radii (panels (d)-(f)) for $^{3}$H, $^{3}$He and $^{4}$He at
    different values of the regulator-cutoff $\Lambda$. All results
    are obtained using IR extrapolations of NCSM results for
    oscillator frequencies $\hbar \Omega \in[35,75]$ MeV in 61(25)
    major oscillator shells for $A=3(4)$ systems, respectively. The
    shaded bands indicate the order of magnitude in the uncertainties
    due to subleading IR corrections. Note the plateaus -- indicating
    RG invariance -- with MPWC and the apparent lack thereof for
    energy results (panels (a)-(c)) based on the WPC interactions, as
    manifested by the sharp jump around $\Lambda \approx 1000$
    MeV. See the text for details.}
\label{fig:ncsm-ir}
\end{figure*}

In this Section we present the results from NCSM few-nucleon
calculations of the bulk properties of $^{3}$H and $^{3,4}$He based on
the $\chi$EFT potentials in MWPC at LO and NLO presented above. For
these calculations we employed the MWPC(40) LO potential with the
$\tilde{C}_{^3P_2}$ LEC calibrated to reproduce the Nijmegen phase
shifts at relative momentum $k \sim m_{\pi}$ ($\Tlab=40$ MeV). For
comparison we also performed NCSM calculations using the MWPC(200) LO potential
where the $^{3}P_{2}-^{3}F_{2}$ channel was renormalized at
$\Tlab=$200 MeV. However, since the two renormalized $^3P_2-^3F_2$ partial-wave contributions are both small at lower energies (e.g., $T_{lab}<100$ MeV), 
 these two different strategies for calibrating the
$\tilde{C}_{^3P_2}$ LEC produce at most $5\%$ relative differences in
the energies and radii for $A=3,4$ nuclei.

Ground-state energies at LO and NLO, and radii at LO, are obtained
using the translationally invariant Jacobi-NCSM
method~\cite{Navratil00} in a harmonic-oscillator basis. For all
calculations we employ rather large oscillator frequencies $\hbar
\Omega$ such that we can capture the high-energy components of the
potential for large values of the ultraviolet regulator cutoff
$\Lambda$. To estimate the values of the model-space converged results
we extrapolate in the infrared (IR) momentum
scale~\cite{ir,ir0,ir2,ir3,ir4} using the formalism outlined in
Ref.~\cite{ir1}. All extrapolations are based on a set of NCSM
calculations carried out for $\hbar \Omega \in[35,75]$ MeV using
61(25) major oscillator shells for $A=3(4)$, respectively. The
extrapolation approach~\cite{ir1} allows an order of magnitude estimate of the
magnitude of subleading IR corrections, which we will indicate with an
uncertainty band.

The MWPC potentials are known to generate RG-invariant \nn{}
amplitudes. Thus, we expect the numerical values for each observable
in our NCSM calculations to exhibit a plateau with respect to large
values of the regulator-cutoff $\Lambda$. Should this plateau not
manifest itself, it would be a clear signature of missing counterterms
necessary to absorb the exposed short-range physics at the present
order. Such a deficit is clearly visible in, e.g., LO predictions of
the ground-state energies in $A=3,4$ systems when using WPC, as shown
in the left panels of Fig.~\ref{fig:ncsm-ir}. For those calculations,
the energies exhibit a clearly noticeable jump at $\Lambda \approx 1000$ MeV indicating a possible divergence. This
behavior is due to the well-known inconsistency in WPC that originates
in the lack of necessary $P$-wave counterterms~\cite{nogga}. Such
artifacts are remedied in MWPC, and we find that the binding energies
of $^{3}$H and $^{3}$He indeed exhibit convincing signs of plateaus as
$\Lambda \gtrsim 800$ MeV at LO and NLO, see the left panels in
Fig.~\ref{fig:ncsm-ir}. This is in accordance with the known results
of the Faddeev calculations presented in Refs.~\cite{nogga,song},
where the cutoff could also be taken much larger. We have verified
that our LO (NLO) results agree with~\citet{song} (\cite{PhysRevC.100.019901}) when the same \nn{} input and cutoff are
used.
It is challenging to converge Jacobi-NCSM calculations for $A=3$
nuclei when using $\Lambda \gtrsim 1.2$ GeV. Using large oscillator
frequencies we observe increasing extrapolation uncertainties due to
subleading IR corrections.

For $^{4}$He, the model-space convergence of the NCSM calculations,
using interactions with larger cutoffs, are associated with larger
uncertainties in the IR extrapolation, see bottom row of panels in
Fig.~\ref{fig:ncsm-ir}. Still, we see the first signs of an
RG-invariant description of $^{4}$He in MWPC.

We note that the error bands presented in this work do not include any
estimate of the order-by-order EFT truncation error. Here, we focus on
the prerequisites, i.e. RG invariance, for enabling an EFT-based
analysis of the epistemic uncertainty. Nevertheless, the
cutoff-variation of the results presented in this work serve as a
rough handle on the truncation error.  More detailed discussions
regarding this subject can be found in
Refs.~\cite{harald,harald2,Wesolowski_2016,bay2,bay3,bay4,bay5,bay6,bay7}.

For point-proton radii, we make predictions at LO, as shown in the
right panels of Fig.~\ref{fig:ncsm-ir}. Again, IR extrapolations were
employed following Ref.~\cite{ir1}. The bands indicate rather larger
uncertainties from subleading IR corrections, which is consistent with
the need to employ large oscillator frequencies in the
NCSM. Nevertheless, we observe a similar plateau for radii as for the
energies, and claim to observe the first signs of RG-invariant
predictions in this observable.

The MWPC results agree rather well with the experimental values, and
the size of subleading corrections to the ground-state energy (as seen
when going from LO to NLO) is very promising. In fact, up to and
including NLO, where additional $S-$wave physics is included, the
energies in $^{3}$H and $^{3}$He reproduce experiment nearly exactly,
which also indicates that higher-order contributions should be rather
small. The impact of such corrections remains to be explicitly
tested. For $^{4}$He, the higher-order contributions must be slightly
larger, which is also expected already in dimensional
counting~\cite{Friar:1997wm}. Overall, MWPC appears to make realistic
energy/radius predictions for few-nucleon systems with mass-numbers $A
\leq 4$.

\subsection{Perturbative calculations in the NCSM}
All LO calculations were carried out in a fully non-perturbative
fashion, while the NLO results in MWPC were obtained
perturbatively. In practice, using the Jacobi-coordinate NCSM code we
obtained the NLO results presented above using a procedure based on the
Hellmann-Feynman theorem. First, we multiply the NLO interaction
potential in Eq.~(\ref{nlomwpc}) with a small coefficient and
subsequently solve the 3- and 4-body Schr\"odinger equations
non-perturbatively. By examining the results as a function of the
small coefficient, the perturbative contribution can be reliably
extracted. See, e.g., Section IV A in Ref.~\cite{trap} for the detailed
procedure.

It is also possible to directly evaluate the expectation value of
first-order perturbation theory. In the NCSM this can be done with
minimal modifications by terminating the iterative Lanczos
diagonalization after a single matrix-vector multiplication using the
LO eigenstate as pivot vector. We implemented this approach in the
M-scheme code \texttt{pANTOINE}~\cite{ir1}, and verified that the
different procedures agree for the $A=3,4$ results.
The ability to perform this kind of extraction is a crucial step
toward the implementation of any perturbative scheme. Starting from
second order in perturbation theory, it is more involved to directly
evaluate the perturbative corrections. On the other hand, the
Hellmann-Feynman procedure can be carried out to extract the
perturbative contribution at arbitrary order without much modification
of current NCSM codes.

\section{Predictions for $^{6}$Li and $^{16}$O using $\chi$EFT
  potentials in MWPC\label{sec:manybodypred}} In this Section we
present NCSM and CC predictions for the ground-state energies of
$^{6}$Li and $^{16}$O at LO and NLO using MWPC. For $^{6}$Li we also
compute the point-proton radius and the ground-state quadrupole moment
at LO. Potentials based on
RG-invariant formulations of $\chi$EFT, e.g. MWPC, have never been
employed for predicting nuclei in the $p$-shell or beyond. Our main
focus here is to study the evolution of the ground-state energy in
selected $A>4$ nuclei as we increase the regulator cutoff
$\Lambda$. The enlargement of the cutoff leads to an enhanced ultraviolet
part of the potential. In the NCSM this ultraviolet physics must be
captured by enlarging the model space, which induces an exponential
increase in basis size. We find that it becomes very challenging to converge
the ground-state energy and wavefunction for $A>4$ nuclei with
$\Lambda \gtrsim 600$ MeV. The ultraviolet component also causes
difficulties in producing a reasonable reference state for the CC
calculations. In this work we obtained reliable results for $^6$Li and
$^{16}$O up to $\Lambda \approx 650$ and 600 MeV, respectively.

For nuclei with mass number $A > 4$, the effect of $P$-waves becomes
more relevant. As outlined in Sec.~\ref{sec:MWPC}, we have constructed
LO potentials, labeled MWPC(40) and MWPC(200), where the LEC in the
$^{3}P_{2}$-wave was renormalized to reproduce phase-shift data in two
different ways, see Figs.~\ref{fig:ph_panel}-\ref{fig:3pf_200}. The
NLO potential in MWPC, see Eq.~(\ref{nlomwpc}), only affects the
$^{1}S_{0}$-wave and is identical for MWPC(40) and MWPC(200). For the
MWPC(40) interaction, the phase shifts in the $^{3}P_2-^{3}F_2$
channel are overly attractive. In contrast, the MWPC(200) potential
exhibit more repulsive phase shifts. A detailed study of how the LO
description of the $^{1}S_0$ phase shifts in $\chi$EFT impacts nuclear
ground-state energies is presented in Sec.~\ref{alt_1s0}.
\begin{figure*}[t]
\includegraphics[width=\textwidth]{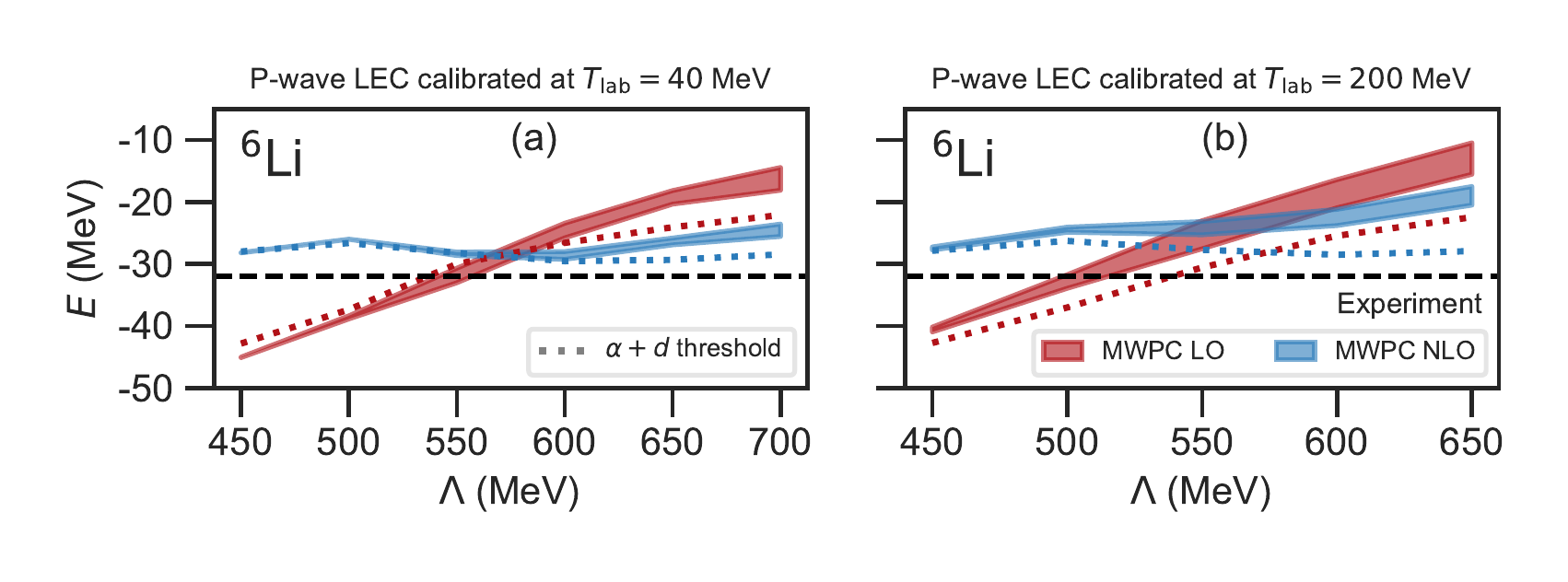}
\caption{Ground-state energies $E$ for $^6$Li at LO and NLO using
  MWPC(40) (panel (a)) and MWPC(200) (panel (b)) as a function of the
  regulator cutoff $\Lambda$. The MWPC NLO results exhibit a markedly
  weaker dependence on the cutoff $\Lambda$ and lie closer to the
  experimental result (dashed line). The bands indicate the estimated
  uncertainties from subleading IR corrections. The dotted lines show
  the $\alpha+d$ threshold (using consistent interactions).}
\label{eb_li6}
\end{figure*}
\subsection{NCSM calculations of $^{6}$Li in MWPC}
As demonstrated in Sec.~\ref{sec:fewbodypred}, the bulk properties of
few-nucleon systems with mass number $A\leq4$ can be described
reasonably well using MWPC. Furthermore, the results exhibit
signatures of RG-invariance which is a minimal requirement of an
EFT. In some ways, $^6$Li constitutes the simplest nucleus beyond
$^{4}$He. It consists of only two more nucleons, and with the
additional proton and neutron naively represented as harmonic
oscillator $P-$wave single-particle states. Here, we perform NCSM
calculations of $^6$Li using the M-scheme code
\texttt{pANTOINE}~\cite{ir1} with oscillator basis frequencies $\hbar
\Omega \in[30,55]$ MeV in 20 major oscillator shells ($N_{\rm
  max}=18$). The relatively large frequencies are needed to improve
the ultraviolet convergence for higher values of the regulator
cutoff. We study regulator cutoffs $\Lambda \in [450,700]$~MeV and
$\Lambda \in [450,650]$~MeV, in 50 MeV increments, for MWPC(40) and
MWPC(200) interactions, respectively. Consequently, we again adopted
the infrared extrapolation scheme from Ref.~\cite{ir1}. It should also
be noted that the NLO corrections were computed perturbatively with
\texttt{pANTOINE} for $^6$Li.

The effects of relative $P$-waves on the ground-state energy of $^6$Li
is obvious when comparing the results for MWPC(40) and MWPC(200) in
Fig.~\ref{eb_li6}. 
Full convergence with respect to $\Lambda$ is not reached for $^6$Li due to the computational limitations. However, our
results indicate that the ground-state 
at LO is less bound than $^4$He plus $^2$H ($\alpha+d$ threshold)
obtained with the same interaction once  $\Lambda \gtrapprox 550$. This is a signature that MWPC does
not generate a physical description of the $^{6}$Li state, which
should be bound with respect to the $\alpha+d$ threshold by nearly 2
MeV. We note that this unphysical behavior has been observed also with
WPC at LO~\cite{Binder:2018pgl}.  Furthermore, this unphysical
description appears to persist at NLO. However, some care is needed
when interpreting our results. The NCSM method includes all
particle-hole excitations in the model space, and within IR
uncertainties, the ground-state energy should at least reside on the
threshold.
The difference between the envelope of the IR uncertainty band and the
threshold indicates that the extrapolation error is
underestimated---at least for larger values of the cutoff
$\Lambda$. We also note that the ground-state energies obtained with
MWPC(40) resides below the $\alpha+d$ threshold for $\Lambda \leq 550$
MeV, and cross the threshold at $~\sim 600$ MeV, where we still
consider our NCSM results to be reasonably well-converged.  In fact,
for both MWPC(40) and MWPC(200), the decreasing rate of $^{6}$Li
binding against cutoff at LO appears to be linear (with fixed slopes)
before and right after crossing the threshold.  Thus, one could not
infer any obvious shift in the wavefunctions---which would be a
signature of a sudden change in the pole structure.  As a result,
rather than immediately concluding that something is fundamentally
wrong on the PC side, we cannot rule out that the apparent failure of
MWPC is simply an effect of fine-tuning in the LECs. Nevertheless, the
predicted LO ground-state energy for $\Lambda\geq600$ ($E\lesssim-15$
MeV) is far from the experimental value $-32$ MeV. This
strong underbinding implies that the effect of higher orders in MWPC
must be sizeable, and this points to the possible need
for some modification of the LO potential in MWPC.

\begin{figure*}[t]
\includegraphics[width=\textwidth]{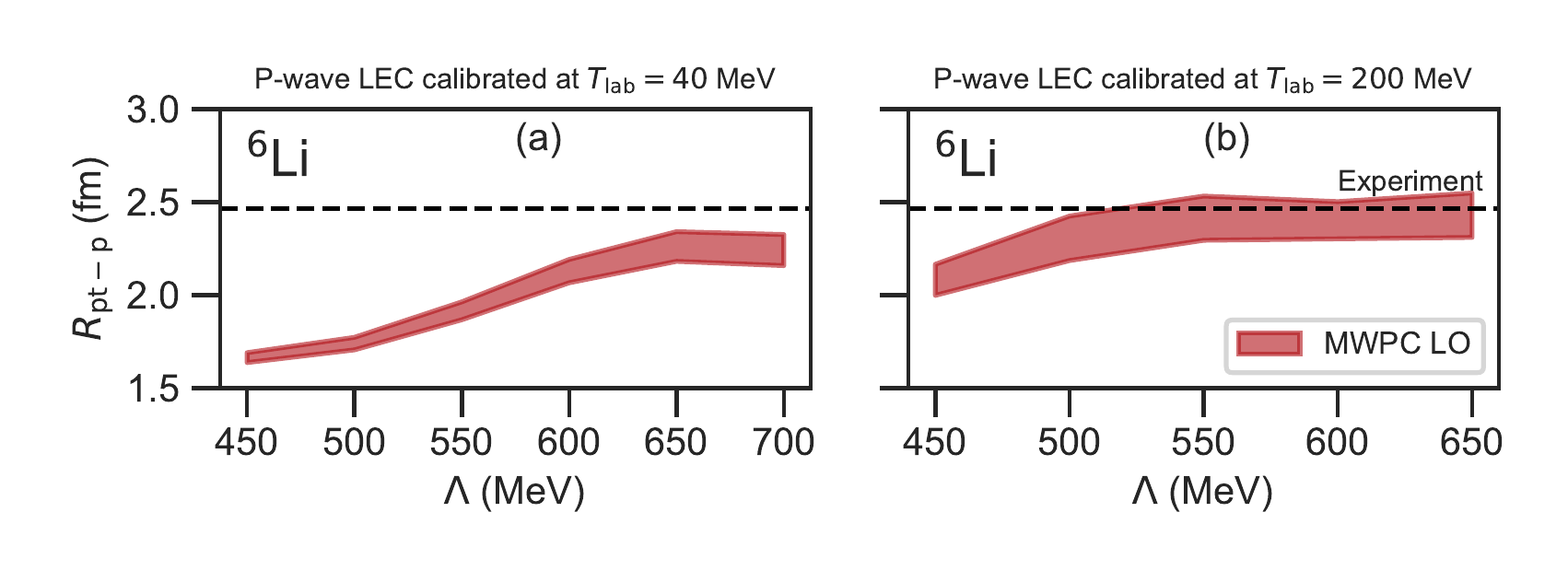}
\caption{Point-proton radii $R_{pt-p}$ for $^6$Li at LO using MWPC(40)
  (panel (a)) and MWPC(200) (panel (b)) as a function of the
  regulator cutoff $\Lambda$. The bands indicate the estimated
  uncertainties from subleading IR corrections. The dashed line marks
  the point-proton radius corresponding to the experimentally
  determined charge radius.}
\label{r_li6}
\end{figure*}

Given our model-space restrictions, and the consequent use of large
oscillator frequencies, it is challenging to reach converged
predictions for the point-proton radius of $^{6}$Li. The estimated
uncertainties coming from the IR extrapolation are sizeable, see
Fig.~\ref{r_li6}.
We note that the radius results for $\Lambda\gtrsim 550$ MeV are
unphysical in the sense that we quote a finite radius for an unbound
system. Indeed, for $\Lambda\gtrsim 550$ MeV the obtained ground-state
energy for $^{6}$Li is above the $\alpha+d$ threshold and the NCSM
basis truncation imposes an IR cutoff that limits the radius
prediction. For $\Lambda \leq 550$ MeV, with MWPC(200) the radius is
predicted slightly larger, and also closer to experiment. This
behavior is intuitively consistent with the slightly lower binding
generated by this interaction. Overall, our results indicate that the
predicted radius of $^{6}$Li is too small compared to experiment. This
correspond to a too large central density of $^{6}$Li. If this
persists to other nuclei, it implies a too large saturation density of
nuclear matter, which is a well-known
problem~\cite{Binder:2014fq,nnlosat} in nuclear structure theory that
seems to persist when using MWPC. Recent analyses suggest that this
problem might be resolved by the explicit inclusion of the
$\Delta(1232)$ degree of freedom in
$\chi$EFT~\cite{nnlodelta,PhysRevC.94.064001, weiguang20}.

Finally, we also studied the quadrupole moment of the $^{6}$Li ground
state, which is experimentally known to be very small and
negative~\cite{til02:708} $Q = - 0.0818(17)$ $e \,
\mathrm{fm}^2$. This small value results from a cancellation of wave
function components and is consequently very sensitive to details of
the nuclear structure. The small quadrupole moment has been
successfully reproduced with ab initio NCSM calculations using
phenomenological, realistic
\nn\ interactions~\cite{Forssen:2009vu}. Using the MWPC(200) LO
interaction from this work we find, however, that we obtain a large
positive quadrupole moment for $\Lambda = 450$~MeV, a small one for
$\Lambda = 500$~MeV and a negative one for $\Lambda = 550$~MeV. The
evolution of the predicted quadrupole moment as a function of the
regulator cutoff points in the direction of varying single-particle
structures. This finding will also be verified with the $^{16}$O
results in the next Subsection. We note that a full convergence study
remains to be performed, but that the observed trend is robust with
the respect to changes in the oscillator frequency and the size of the
model space.

\subsection{$^{16}$O in MWPC}
We now turn to the case of the doubly-magic nucleus $^{16}$O and
calculate its ground-state energy using interactions from MWPC at LO
and NLO. The oxygen isotopic chain has been extensively studied with
\textit{ab initio} methods and chiral potentials in
WPC~\cite{PhysRevLett.105.032501,PhysRevLett.108.242501,PhysRevLett.110.242501,nnloopt,PhysRevLett.111.062501,PhysRevLett.112.102501,PhysRevLett.117.052501,nnlosat,PhysRevLett.123.252501}.
These calculations have revealed that an accurate description of
binding energies, radii, and spectra is very sensitive to fine details
of the employed chiral potential model. Furthermore, in
Ref.~\cite{bay4} it was found that simultaneously optimized chiral
\nn{} and \nnn{} interactions from WPC at NNLO predicts $^{16}$O to be
unbound with respect to decay into four $\alpha$-particles.
Interestingly, recent calculations~\cite{nnlodelta, weiguang20} based
on chiral potentials with explicit inclusion of $\Delta$-isobars at
NNLO found $^{16}$O to be bound. This result might also indicate an
important role of the finite nucleon-size for reproducing saturation
properties in nuclei.

We will use single-reference \textit{ab initio} CC theory to calculate
the ground-state of $^{16}$O. The many-nucleon wavefunction is
represented via an exponential ansatz $\vert \Psi \rangle = e^T \vert
\Phi_0 \rangle$, where $\vert \Phi_0 \rangle$ is an uncorrelated
reference state commonly chosen as the Hartree-Fock (HF)
ground-state. Many-body correlations are then included by acting with
$e^T$ on the reference state, where $T= T_1 + T_2 \ldots$ is a linear
expansion in particle-hole excitations typically truncated at some low
excitation rank. In this work we truncate $T$ at the singles-doubles
excitation level. When using spherical CC, we also include triples
excitations perturbatively in an approach known as the
$\Lambda$-CCSD(T)
approximation~\cite{taube2008,Hagen:2010gd,hagen2014}. For
closed-shell systems that can be well described using a
single-reference formalism, this approximation has been shown to
account for about 99\% of the full correlation energy~\cite{cc5}. We
remind the reader that the CC method is non-variational, and as a
consequence the Hellmann-Feynman theorem is strictly not valid when
evaluating expectation values when the cluster operator $T$ is
truncated (see e.g. ~\cite{Hagen:2010gd} for more details). We
therefore compute the perturbative corrections at NLO as an
expectation value using the LO CC wavefunction.

\begin{figure}[t]
  \includegraphics[width=\columnwidth]{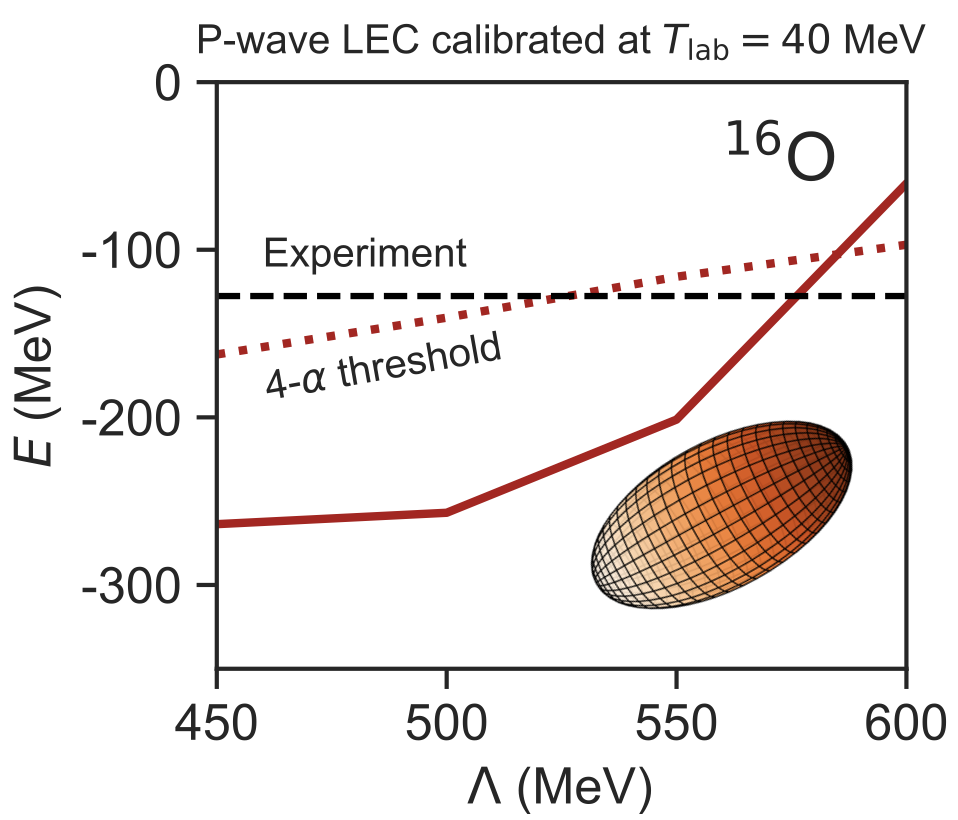}
\caption{The ground-state energy of $^{16}$O versus regulator cutoff
  $\Lambda$ at LO in MWPC(40). The ground state is axially deformed
  for all values of the cutoff in this figure.}
\label{fig:16O_mwpc}
\end{figure}

With MWPC(40) at LO we find that the Hartree-Fock (HF) single-particle
orbitals exhibit an unconventional ordering with a $1d_{5/2}$ orbital
below the $1p_{1/2}$ orbital and a very large ($\geq 80$ MeV)
splitting between the $1p_{1/2}$ and $1p_{3/2}$ orbitals. Although
these single-particle orbitals are not observable
quantities~\cite{duguet2012}, the observed ordering is in stark
contrast with the traditional single-particle shell-model picture of
~\citet{mayer1949}, which usually provides a realistic starting point
for describing well bound nuclei near the valley of
beta-stability. This untraditional ordering is presumably caused by the
over-attractive $^1S_0$ and $^3P_2$ partial-waves, as shown in
Figs.~\ref{fig:ph1s0} and \ref{fig:ph_panel}. Furthermore, the
inversion of the $d_{5/2}$ and $p_{1/2}$ orbitals prevents us from a
spherical single-reference CC description of the ground-state of
$^{16}$O. To compute the ground-state of $^{16}$O using MWPC(40) we
therefore performed CC calculations starting from an axially deformed
Hartree-Fock reference state. Here the Hartree-Fock reference state
was constructed assuming prolate deformation, see ~\cite{novario2020}
for more details.
The ground-state energies are plotted as a
function of the cutoff $\Lambda$ in Fig.~\ref{fig:16O_mwpc}. Being a
doubly-magic nucleus, $^{16}$O should be spherical in its ground
state. Thus we conclude that the MWPC(40) LO interaction is highly
unphysical. Also, at lower cutoffs, the LO result yields a
ground-state energy that is two orders of magnitude from the
experimental value. The unphysical LO results do not motivate a
further study of the NLO corrections.

We performed the calculations using model-spaces sizes of up to 11
major oscillator shells ($N_{\rm max}=10$) and varied the oscillator
frequencies over a wide range ($\hbar\Omega \in[35,60]$ MeV). This
allowed us to find the energy minimum for a given model-space, and
extract reasonably well converged ground-state energies. For example,
for the hardest interaction, $\Lambda=600$ MeV, the CCSD energy at the
$\hbar\Omega $ minimum goes from -58 MeV to -61 MeV when increasing
from $N_{\rm max}=8$ to $N_{\rm max}=10$.
We note that at $\Lambda=600$ MeV, $^{16}$O becomes unstable against
decay into four-$\alpha$ particles. A CC prediction for the
ground-state of $^{16}$O that is unbound with respect to the
four-$\alpha$ decay-threshold requires some care in its
interpretation. Clearly, such an energy does not represent the true
ground-state of the system. Indeed, four $\alpha$-particles very far
apart would yield an energy equal to the corresponding threshold
value. But this very exotic $\alpha$-cluster configuration can not be
described in the CC approach that we use. In order to
describe a state that is dominated by clusterization into
$\alpha$-particles, one would need to include at least $4p$-$4h$
excitations in the cluster amplitude $T$. Such an approach is
currently not possible due to the orders of magnitude increase in
computational cost. 

We now move on to the MWPC(200) interaction, which has a more
repulsive $^3P_2$ component with a better overall agreement to the
Nijmegen phase shifts analysis. We note that there is still an
over-attraction in the NN $^1S_0$ channel as shown in
Fig.~\ref{fig:ph1s0}. With MWPC(200), we find a conventional ordering of
single-particle states. All CC calculations indicate that the
spherical states of $^{16}$O are always more bound than their
corresponding deformed counterparts throughout $\Lambda=450-600$ MeV
for this interaction. However, some pathological behaviors are still
present. Most importantly, we find that for cutoff values
$\Lambda > 450$ the ground-state in $^{16}$O is always unstable
with respect to decay into four $\alpha$ particles. Our spherical CC
calculations were carried out in a model-space up to 17 major
oscillator shells ($N_{\rm max} =16$). 
The results are very similar to those plotted later in the left panel of Fig.~\ref{fig:16O_p} (i.e., MWPC(200) with perturbative P-waves).

\begin{figure}[t]
\includegraphics[width=\columnwidth]{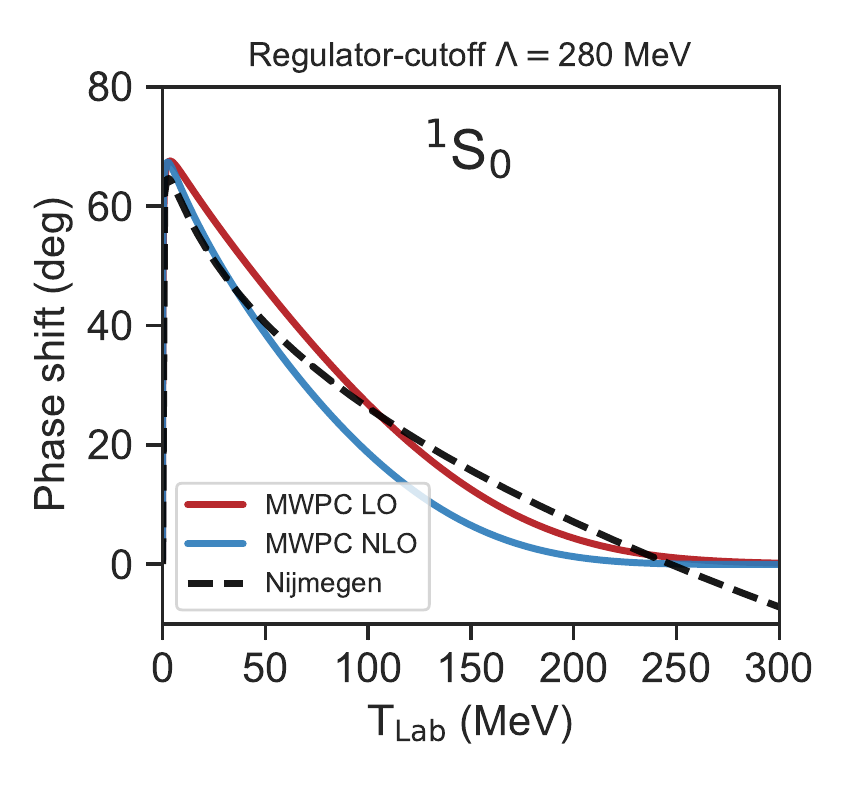}
\caption{MWPC LO and NLO $^{1}S_{0}$ phase shifts using a
  regulator-cutoff $\Lambda=280$ MeV as a function of laboratory
  scattering energy $\Tlab$. The MWPC LO interaction yields slightly
  more attractive phase shifts.}
\label{ph_1s0280}
\end{figure}

There are most likely several possible origins that contribute to the
failures of MWPC(40) and MWPC(200) in producing a physical $^{16}$O
ground-state. First, as already seen in the case of $^6$Li, the
effects due to different strategies for calibrating the LO LEC in the
$^3P_2$-$^3F_2$ partial-waves are further magnified in $^{16}$O. For
example, at $\Lambda=450$ MeV, MWPC(40) and MWPC(200) yield vastly
different shapes and energies for the ground state; $-264$ MeV
(deformed) and -150 MeV (spherical), respectively. From the results
presented above, we have to conclude that MWPC cannot be employed for
realistic predictions of atomic nuclei beyond $^{4}$He. However, we
would like to point out that it is possible to obtain a remarkably
good descriptions of the ground-state energies of $^{4}$He as well as
$^{16}$O at LO in MWPC if one tunes the regulator cutoff $\Lambda=280$
MeV. At this value, the LO description of the $^{1}S_{0}$ phase shift
is qualitatively very similar to Nijmegen data and the NLO correction
is small, see Fig.~\ref{ph_1s0280}. This particular LO interaction
yields $^{16}$O and $^{4}$He binding energies 127.3 MeV and 29.5 MeV,
respectively. The $^{16}$O ground state is also spherical.
Using the regulator cutoff $\Lambda$ as one of the fitting parameters
in this way eliminates the model-independent aspects of $\chi$EFT and its
fundamental connection with QCD will be lost. Of course, an
interaction with a tuned regulator cutoff could still be useful for
guiding experiment and phenomenology. However, in a $\chi$EFT with a
PC that yields RG-invariant observables, no particular cutoff is
preferred. One can choose any value $\Lambda \gtrsim \Lambda_b$ for a
quantitative calculation, and then try to estimate residual,
i.e. higher-order, cutoff dependencies and EFT
errors~\cite{harald,harald2}. Setting $\Lambda=280$ MeV, which is most
likely much smaller than the breakdown scale $\Lambda_b$ of $\chi$EFT,
removes a large chunk of the relevant low-momentum dynamics from the
loops. As such, this model will probably extrapolate unreliably to
larger mass-numbers $A$ and it will be difficult to assign a
physics-based EFT uncertainty to the results.

\section{Predictions based on other RG-invariant power counting schemes\label{sec:otherpcs}}
\label{alt_1s0}
The results for $^{6}$Li and $^{16}$O presented in the
previous sections most likely rule out the usefulness of MWPC for heavier
systems. Obviously, one needs to seek for alternative PCs. In fact,
the large discrepancy between the Nijmegen analysis and the LO
$^{1}S_{0}$ phase shift as shown in Fig. \ref{fig:ph1s0} has already
motivated research on several such alternative PCs.

\subsection{The dibaryon field}
Potentials in $\chi$EFT that employ the dibaryon (db) auxiliary
field~\cite{dbk,BY,1s0d} give a very good, and RG-invariant,
description of the $^{1}S_0$ phase shift at LO. However, in such
approaches, the resulting potential contains an energy-dependent
short-range term
\begin{equation}
V_{\rm db}^{\rm LO}(E)=\frac{1}{\Delta +c\cdot E},  \label{db}
\end{equation}
where the on-shell energy $E=k_{0}^{2}/m_{N}$, with $k_{0}$ denoting
the on-shell center-of-mass momentum, and $m_{N}=938.9$ MeV the
nucleon mass. The two parameters $\Delta$ and $c$ are LECs to be renormalized.  Note that
in order to reproduce the amplitude zero at $\Tlab\sim 250$ MeV, the
PC proposed in Ref.~\cite{1s0d} incorporates one more LEC which
effectively has the same structure as the usual $C_{1S0}$ contact
term, i.e.,
\begin{equation}
V_{\rm DBZ}(E)=C_{1S0}+V_{\rm db}^{\rm LO}(E). \label{dbz} 
\end{equation}
We denote this PC as DBZ (dibaryon potential which reproduces the
amplitude zero). The long-range part of the OPE potential is then
iterated non-perturbatively together with the short-range $V_{\rm
  DBZ}$ potential in the Lippmann-Schwinger or Schr\"odinger
equations.

In the \nn{} sector, solutions of the Lippmann-Schwinger or
Schr\"odinger equations based on an energy-dependent potential can be
obtained straightforwardly. The only caveat is that eigenfunctions are
no longer necessarily orthogonal~\cite{edep,edep1}. On the other hand, it is very
difficult to solve a many-nucleon Schr\"odinger equation based on
energy-dependent potentials. In order to proceed, approximations are
needed. We studied the predictions from energy dependent DBZ potentials for
describing $A=3,4$ systems, and tried to quantify the uncertainties due
to non-unique transformations of an energy-dependent potential to a
purely momentum-dependent potential $V(p,p')$. We adopted the
following procedure. First, we solve the two-body Schr\"odinger equation
\begin{equation}
(H_{0}+V(E))|\psi\rangle =E|\psi\rangle ,  \label{s}
\end{equation}
with $H_{0}$ the kinetic energy, $\psi$ the eigenfunction and $V(E)$
the LO potential which contains both momentum- and
energy-dependencies. We represent the total Hamiltonian $H$ in a finite
momentum basis with $N \approx 100$ states that cover momenta
$[0,\Lambda_p]$ and where we also ensure that
$\Lambda_p>\Lambda$. Thereafter, it is straightforward to solve
Eq.~(\ref{s}) for $N$ eigenvectors iteratively until the corresponding
difference between the on-shell energy $E$ on both sides of the
equation falls below a convergence criterion
$\varepsilon_E=10^{-16}$. The resulting self-consistent eigenfunctions
and eigenvalues are denoted as $\psi_{E_i}$ and $E_i$. Due to the
energy dependence, the span of eigenvectors $\{\psi_{E_i}\}_{i=1}^{N}$
does not form an orthogonal basis. Instead, we choose to employ the
Gram-Schmidt orthogonalization method to form an orthogonal basis
$\{\psi^{\rm GS}_{E_i}\}_{i=1}^{N}$. This transformation is not
unique. Indeed, we can start from any of the $N$ vectors in the
Gram-Schmidt procedure and generate a different basis. However, equipped with any orthogonal basis
we can reconstruct an on-shell equivalent Hamiltonian
\begin{eqnarray}
\langle p|H|p^{\prime }\rangle &=&\ooalign{$\textstyle\sum$\cr\hidewidth$
\displaystyle\int$\hidewidth\cr}_{E_{i}}\ooalign{$\textstyle\sum$\cr
\hidewidth$\displaystyle\int$\hidewidth\cr}_{E_{i}^{\prime }}\langle p|\psi^{\rm GS}
_{E_{i}}\rangle \langle \psi^{\rm GS}_{E_{i}}|H|\psi^{\rm GS} _{E_{i}^{\prime }}\rangle
\langle \psi^{\rm GS} _{E_{i}^{\prime }}|p^{\prime }\rangle  \label{e2} \\
&=&\ooalign{$\textstyle\sum$\cr\hidewidth$\displaystyle\int$\hidewidth\cr}
_{E_{i}}\langle p|\psi^{\rm GS} _{E_{i}}\rangle E_{i}\langle \psi^{\rm GS} _{E_{i}}|p^{\prime
}\rangle .  \label{e2a}
\end{eqnarray}%
Subtracting the kinetic term yields a momentum-dependent potential which preserves all of the
original eigenvalues%
\begin{equation}
V(p,p^{\prime })=\langle p|H|p^{\prime }\rangle -\frac{p^{2}}{m_{N}}\delta
_{pp^{\prime }},  \label{e3}
\end{equation}
and which we then use in the many-body calculations.
After renormalization, the numerical value of $c$ in
Eq.~(\ref{db}) turn out to be of order $10^{-4}$ smaller than $\Delta $
and $C_{1S0}$\footnote{Here the 3 LECs are fitted to reproduce
  $a_0=-23.7$ fm, $r_0=2.7$ fm, and the Nijmegen phase shift analysis
  at $\Tlab=250$ MeV.}, which likely correspond to a small
non-orthogonality between the vectors $\psi _{E_{i}}$. On the other hand, we found that
without any Gram-Schmidt re-orthogonalization the resulting
$V(p,p^{\prime })$ potential will generate phase shifts which deviate
about $15\%$ from the original values, i.e. the ones given by
$V(E)$. The non-uniqueness of the re-orthogonalization procedure is
manifested in the momentum-dependent potential as off-shell
modifications. The size of this effect can be explored by selecting
different $\psi _{E_{i}}$ as the initial vector in the Gram-Schmidt
procedure to generate potentials $V(p,p^{\prime })$ with differing
off-shell behavior. We have carried out this test, and find that this
effect is propagated into many-body calculations and gives about
$10\%$ ($20\%)$ variation in the ground-state energies of $^{3}$H
($^{4}$He). Besides this variation due to the energy dependence of the
potential, we find that the reproduction of the corresponding ground-state energies
is comparable to the MWPC result, i.e. slight underbinding. The main
difference is that the NLO correction to the energy appears to be
smaller, which can be expected judging from the NLO correction at the
\nn{} phase-shift level~\cite{1s0d}. We also note that shuffling the
order of eigenvectors in the Gram-Schmidt procedure creates
discontinuities in the first derivative of $V(p,p^{\prime })$ with
respect to $ p$ and $p^{\prime }$. Although this is not forbidden in
principle, it could create artifacts of numerical origin in many-body
calculations.

The evaluation of DBZ at NLO is even more involved compared to LO, as the
NLO interactions again contain additional energy dependencies, which
read
\begin{equation}
C_{2}^{\rm DBZ}+D_{2}^{\rm DBZ}E+\alpha V_{\rm DBZ}^{\rm LO}(E)+\beta \lbrack V_{\rm DBZ}^{\rm LO}(E)]^{2},
\label{dbznlo}
\end{equation}%
with $C_{2}^{\rm DBZ}$, $D_{2}^{\rm DBZ}$, $\alpha $ and $\beta $ denoting
four new LECs, and $V_{\rm DBZ}^{\rm LO}$ the short-range LO potential
as defined in Eq.~(\ref{dbz}) (which is not re-fitted at NLO). A
direct perturbative evaluation in the \nn{} sector is straightforward. 
However, the renormalized interaction is energy dependent and
cannot easily be used in many-body calculations. To our knowledge,
there is no strict phase-shift equivalent transformation method to be
applied perturbatively, as in the \nn{}-case. There are several
approximate ways to transform the energy-dependent NLO terms in the
DBZ potential to a purely momentum-dependent representation. The
bottom line in all such methods is to treat the
energy-dependent terms in Eq.~(\ref{dbznlo}) as small perturbations to
the LO amplitude. 
Once all the
energy-dependent terms are transformed (either one-by-one individually or as a whole) into purely momentum-dependent
terms, one must renormalize the LECs associated with those
transformed (i.e., momentum-dependent) contact terms at NLO. One could
also test the possibility of approximating $E$ by $\frac{p^2+p'^2}{m_N}$ for the second term of
Eq.~(\ref{dbznlo}), inspired by the equation of motion. In all our attempts to transform the energy
dependence to a pure momentum dependence, the four NLO LECs always
yield phase shift equivalent results, as expected. The off-shell
differences, however, manifest themselves in an uncontrollable fashion
in many-body calculations. Our analysis indicates that the
approximate, and non-unique, transformation of the energy-dependent
DBZ potential at LO yield sizeable errors that increase with
mass-number $A$. We do not present any detailed CC results for
$^{16}$O. We only summarize that in all our calculations based on the
DBZ potential at LO and NLO, we never recovered a $^{16}$O nucleus
that was bound with respect to four-$\alpha$ decay for any values of
the regulator cutoff $\Lambda \geq 500$ MeV.

\subsection{A separable potential}
\begin{figure}[t]
  \includegraphics[width=\columnwidth]{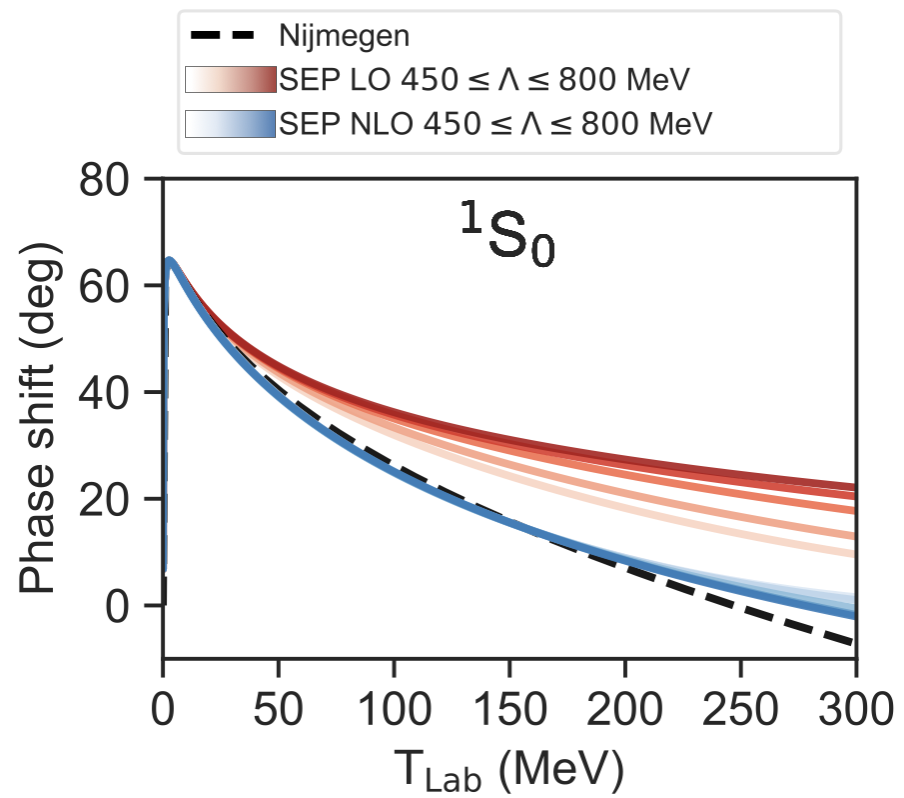}
\caption{LO and NLO $^{1}S_{0}$ phase shifts obtained using the power
  counting based on a separable potential (SEP) defined in
  Ref.\protect~\cite{sep} as a function of laboratory scattering
  energy $\Tlab $. Darker line colors correspond to larger values of
  the cutoff $\Lambda$. The SEP LO interaction exhibits a stronger
  cutoff-variation in the $^{1}S_0$ phase shift compared to the SEP
  NLO interaction. The latter interaction best reproduces the Nijmegen
  result (dashed line). The LO LECs are renormalized to the scattering
  length $a_0=-23.7$ fm and the effective range $r_0=2.7$ fm. The NLO
  LECs are renormalized to the scattering length $a_0=-23.7$ fm, the
  effective range $r_0=2.7$ fm and a best fit to Nijmegen phase shift
  up to $\Tlab=200$ MeV.}
\label{plot_sep_best}
\end{figure}

In an effort to eliminate the energy dependence of the dibaryon field,
while trying to maintain the good reproduction of the $^{1}S_{0}$
phase shift, one could transform the dibaryon structure in the
Lagrangian to yield an energy-independent and separable potential
(SEP)~\cite{Yamaguchi1954}. The LO short-range structure of this
potential reads~\cite{sep}
\begin{equation}
  V_{\rm SEP}(p,p')=\frac{ym_{N}}{\sqrt{p^{2}+m_{N}\Delta}\sqrt{p'^{2}+m_{N}\Delta }}.
  \label{sep}
\end{equation}
This reproduces the short-range physics of one dibaryon field. There
are two LECs, $y$ and $\Delta $, at LO. The full LO SEP potential
contains the above short-range part plus the Yukawa potential. The
resulting $^1S_0$ phase shifts at LO and NLO are given in
Fig. \ref{plot_sep_best}.

To generate the NLO amplitude, one perturbatively inserts the
following NLO short-range terms in the $^{1}S_{0}$ channel
\begin{equation}
  C+y^{(1)}V_{\rm SEP}+\Delta
  ^{(1)}\left(\frac{1}{p^{2}+m_{N}\Delta}+\frac{1}{p^{\prime 2}+m_{N}\Delta
  }\right)V_{\rm SEP},
\end{equation}
where $C,y^{(1)},\Delta ^{(1)}$ are additional LECs, renormalized to
provide $a_0=-23.7$ fm, $r_0=2.7$ fm and a best fit to Nijmegen phase
shift up to $\Tlab=200$ MeV. Note that $V_{\rm SEP}$ here is the
short-range potential already renormalized at LO. The LECs within
$V_{\rm SEP}$ are not re-fitted at NLO. The resulting NLO phase shifts
reproduce the Nijmegen $^{1}S_{0}$ phase shifts quite well for a wide
range of cutoff values $\Lambda$ as listed in
Fig. \ref{plot_sep_best}, which also shows a more reasonable LO to NLO
change comparing to MWPC.

As for MWPC and DBZ, the SEP potential yields a reasonable LO
prediction of the ground-state energies of $^{3}$H, $^{3}$He and
$^{4}$He, with NLO corrections of expected sizes. In fact, this seems
to be a trend; most RG-invariant PCs yield LO and NLO potentials in
$\chi$EFT capable of describing $A\leq 4$ nuclei rather well. As for
DBZ, the small NLO correction to the ground-state energy using the SEP
PC can be inferred from the fact that the space for improvement is
tiny since the LO results reproduce \nn{} phase-shifts in low partial
waves quite well.  Note that our results for $A=3,4$ systems at lower
cutoffs are in agreement with a recent calculation~\cite{mario20},
where the LO treatment in the $^1S_0$ channel is equivalent to our SEP
potential plus one constant contact term.

Having removed the energy-dependence, via the separable formulation,
we used the SEP potential in \textit{ab initio} CC calculations
to predict the ground-state energy of $^{16}$O at LO and NLO for
cutoff values $\Lambda \leq 600$ MeV. We refer to the potentials
associated with the LEC in $^3P_2-^3F_2$ channels fitted up to
$\Tlab=40$ and $\Tlab=200$ MeV as SEP(40) and
SEP(200), respectively.

For SEP(40), we found that the over-attractive $^{3}P_{2}$
partial-wave still generates an ordering of the HF single-particle
states that is in stark contrast to traditional shell model
interpretations, or gives a very large splitting between the $1p_{1/2}$ and
$1p_{3/2}$ states. On these grounds, we discard further analyses of
the SEP(40) interaction. For SEP(200), the single-particle states
exhibit a conventional ordering, which suggests a spherical ground
state. However, we are not able to obtain ground-state energies of
$^{16}$O below the corresponding four-$\alpha$ threshold throughout the
cutoff range $\Lambda=450-600$ MeV. Although this renders the NLO
correction less meaningful, we note that they are always repulsive---which
makes the results even more unphysical.

\begin{figure*}[t]
\includegraphics[width=\textwidth]{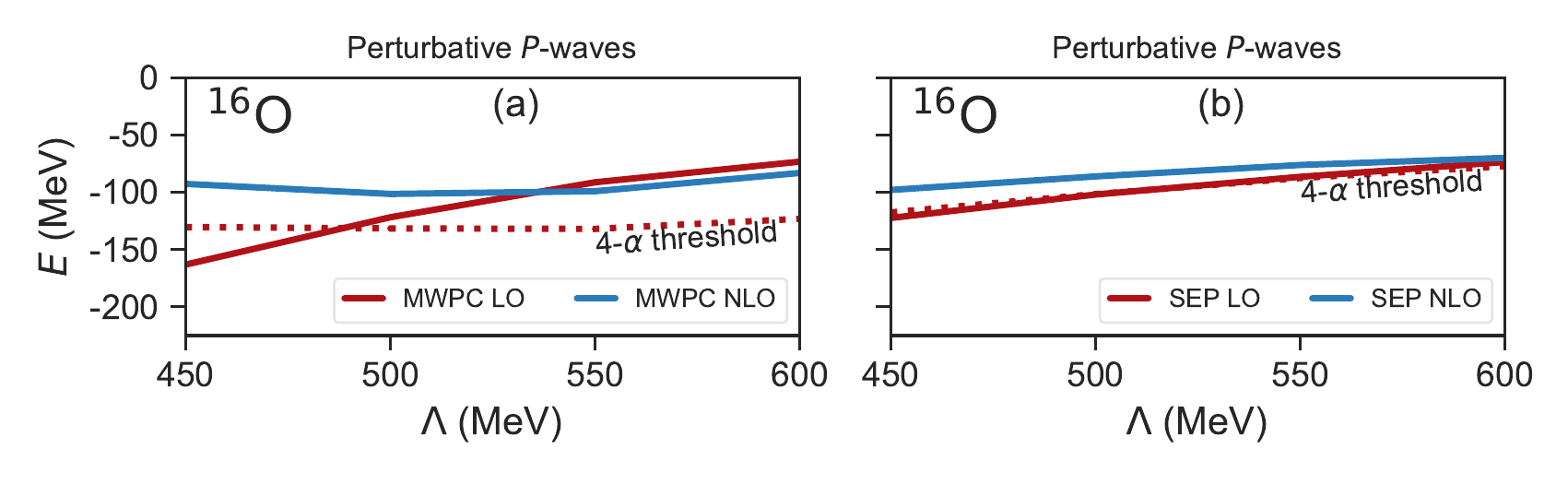}
\caption{Ground-state energy of $^{16}$O versus regulator cutoff
  $\Lambda$ with perturbative $P$-waves at LO and NLO using MWPC
  (panel (a)) and SEP PC (panel (b)). The respective LO four-$\alpha$
  thresholds are indicated by the red dotted lines. The NLO results in
  both panels exhibit a weaker dependence with respect to variations
  in the cutoff $\Lambda$. }
\label{fig:16O_p}
\end{figure*}

\subsection{Perturbative $P$-waves\label{sec:perp}}
It was shown recently~\cite{bingwei18} that all of the $P$-wave
amplitudes, with the exception of $^3P_0$, can be reproduced rather
well in an order-by-order perturbative approach. This particular
$P$-wave still requires a non-perturbative treatment\footnote{For the
  $^3P_0$ channel, a further study~\cite{Peng:2020nyz} suggests a
  perturbative treatment is possible if an additional
  counterterm are promoted to LO in addition to the long-range OPE in this channel.}. Thus, it is possible that the LO
amplitude in $\chi$EFT should only comprise the $^1S_0$,
$^{3}S_{1}-^{3}D_{1}$, and $^3P_0$ waves. To explore the consequences
of this recent PC, we have performed many-body calculations for the
ground-state energies in $^3$H, $^{3}$He, and $^{4}$He using the
Jacobi-NCSM method and $^{16}$O using the CC method based on MWPC and
SEP potentials with perturbative $P$-waves. The LEC in the $^3P_0$
channel is renormalized to Nijmegen phase shifts up to $\Tlab=40$ MeV
as before, and the interaction in all remaining $P$-waves (including
channels coupled to $^3P_2$) vanish identically at LO. Since there can
no longer be any differences in the $^3P_2-^3F_2$ channels due to
alternative calibration procedures, we can drop the (40) and (200)
labels from such interactions.

With perturbative $P$-waves, the $^{3}$H and $^{3,4}$He
binding energies reside between the MWPC(40) and MWPC(200), or SEP(40)
and SEP(200) if the SEP PC is adopted in the $^1S_0$ channel. We also
observe similar convergence patterns as before when going from LO to
NLO. This is expected since nuclei with mass number $A \leq 4$ are
quite insensitive to $P-$waves. Indeed, all our previous results based
on the two different $^3P_2-^3F_2$ calibrations differed at most 5$\%$
for the cutoff values considered in this work.

Turning to $^{16}$O, we found that the Hartree-Fock solutions starting
from MWPC and SEP PC with perturbative $P$-waves give single-particle
states with ordering that allows for a spherical single-reference
CC description. We therefore computed the ground-state
energies of $^{16}$O using the spherical $\Lambda$-CCSD(T)
approximation and the results are presented in
Fig.~\ref{fig:16O_p}. We employed 17 major oscillator shells ($N_{\rm
  max}=16$) and extract a minimum CC energy for $\hbar\Omega
\in[16,50]$ MeV for all values of the cutoff $\Lambda=450-600$ MeV. As
one can see, MWPC with perturbative $P$-waves yields a $^{16}$O ground
state that decays into four $\alpha$-particles for $\Lambda>500$
MeV. For SEP with perturbative $P$-waves, the CC results for the
ground-state energy borders the four-$\alpha$ threshold. Looking
carefully, we find that the ground-state of $^{16}$O becomes unbound
with respect to four-$\alpha$ decay starting at cutoff values
$\Lambda\geq 550$ MeV. However, without a comprehensive uncertainty
analysis and the inclusion of higher-order particle-hole excitations
in the CC method we cannot conclusively determine whether
the SEP yields a stable $^{16}$O ground-state. We do note $\sim 5\%$
increase in the binding energies when employing e.g. the CCSD(T)
approximation instead. This small shift makes $^{16}$O bound with
respect to four-$\alpha$ threshold throughout the cutoff range
$\Lambda=450-600$ MeV.

The NLO correction to the PC with perturbative $P$-waves is not fully
known. It consists of at least the NLO $^1S_0$ contribution in
Eq.~\ref{nlomwpc}, which is what we employ here. In
Ref.~\cite{bingwei18} it is proposed that the long-range OPE
contribution at $^1P_1$, $^3P_1$ and $^3P_2-^3F_2$ channels might
belong to NLO as well, though the relative importance between the NLO
$^1S_0$ contribution and the long-range $P$-wave contributions is yet
to be understood. Unfortunately, even with the less-repulsive choice
--- where only the NLO $^1S_0$ contribution enters --- the NLO shift
is more or less always repulsive for $\Lambda=450-600$ MeV, see
Fig.~\ref{fig:16O_p}. Thus, it appears that the NLO correction to a PC
with perturbative $P-$waves is unlikely to improve upon the results
for $^{16}$O, at least within the cutoff range $\Lambda=450-550$ MeV.
In summary, the success of this PC also appears to be limited to
lighter nuclei.

\section{Summary and implications for future work\label{sec:summary}}
In this work we have performed \textit{ab initio} NCSM and CC
calculations of $^{3}$H, $^{3,4}$He, $^{6}$Li, and $^{16}$O at LO and
NLO in $\chi$EFT using several PC schemes which all exhibit
RG-invariance in the \nn{} sector. We include all subleading
corrections perturbatively. We employed the MWPC of
Ref.~\cite{BY3p0,BYtri,BYs}, the dibaryon fields of Ref.~\cite{1s0d},
the separable PC of Ref.~\cite{sep}, and the recently proposed
perturbative treatment of $P$-waves~\cite{bingwei18}. Our NCSM results
indicate that the predictions of ground-state energies and
point-proton radii in $^{3}$H and $^{3,4}$He, exhibit signatures of
RG-invariance, and the converged values are mainly sensitive to
variations in the two $S$-wave channels. Indeed, using different
strategies for generating the amplitude in the $^{1}S_{0}$
partial-wave, the converged LO binding energies differ by $\sim$1 MeV
in $^{3}$H and $^{3}$He, and $\sim$5 MeV in $^{4}$He. In addition, for the DBZ
approach we find a $\sim$20\% variance due to the non-unique potential
redefinition to handle the energy-dependent dibaryon structure.

The various RG-invariant NLO corrections to the ground-state energies
for $^3$H ($^4$He) are as large as $\sim$2 ($\sim$7) MeV with MWPC
(attractive shifts, i.e., toward the experimental value). The
corresponding shifts are smaller when using the SEP and DBZ
potentials. 
In summary, all of the NLO results for $A=3,4$ nuclei are very
reasonable and certainly agree with experiment at the level expected of NLO calculations, especially 
judging from the fact that no higher-body force has been added yet.
Overall, the various PC schemes yield similar and
realistic descriptions of $A\leq 4$ nuclei.

However, this work has revealed a number of problems when applying the 
RG-invariant PC schemes to the study of nuclei with $A>4$.
Some of the flaws of the MWPC interactions are seen already in the
predictions of the ground-state energy and radius of
$^{6}$Li. Contrary to experiment, the ground state of this nucleus was
predicted slightly above the $\alpha+d$ threshold, even when
considering the estimated uncertainties due to the IR
extrapolation. 
Furthermore, we find too small radii and a strong cutoff dependence in the quadrupole moment.
These results also motivate the detailed study of
$^{16}$O.

Based on the results of CC predictions for the ground-state energy of
$^{16}$O, we conclude that none of the PC schemes employed in this
work appear to yield a realistic description of this nucleus. In fact,
the most important observation is that none of the RG-invariant
PCs in $\chi$EFT successfully manages to generate a realistic and
spherical $^{16}$O ground state at LO, which leaves small hope for a
remedy from perturbative corrections.
Model-space limitations hinder us from explicitly demonstrating
renormalizability at large cutoffs for $A \gtrsim 6$. Still, our
limited-cutoff results reveal fundamental flaws in several of the
recently developed PC schemes, which is a significant and unexpected
discovery.
Of course, future analyses of RG-invariant schemes deserve a more
careful and systematic parameter estimation of the LECs and adequate
handling of the model discrepancy due to neglected higher-order
diagrams in the $\chi$EFT expansion.

In summary, it appears that the essential nuclear-binding mechanism fails in all
present RG-invariant PC schemes for $\chi$EFT. A remedy to this
important finding will be critical for the utilization of PC schemes
that employ a perturbative inclusion of subleading orders. We conclude
that one (or several) of the following scenarios must be true:
\begin{enumerate}
  \item[(i)] We have failed to capture a very fine-tuned process
in the renormalization of the relevant LO LECs that is responsible for
generating realistic ground states in $^{16}$O and $^{6}$Li.
  \item[(ii)] There is a scale critical to the physical description of finite-size nuclei, which is not captured by the
contact terms at $\Lambda\rightarrow \infty$. This conceivable scale was discussed
recently in Ref.~\cite{pionlessfail} using pionless EFT, although the possible
implications in $\chi$EFT remain unclear.
  \item[(iii)] Something else
is missing in the LO interaction for describing $^{16}$O and $^{6}$Li,
and most likely other nuclei.
\end{enumerate}
Naturally, we cannot rule out scenarios (i) and (ii), but we would
like to speculate that it is quite possible that, due to an increasing
relative importance of many-body forces in larger systems---as sketched
also in Section 4.3 of Ref.~\cite{yang20}---a \nnn{} force, such as
the $\Delta$-full Fujita-Miyazawa NNN force\cite{Fujita1957} or the
$\Delta$-less NNN force, must be promoted to LO in a $\chi$EFT for
many-nucleon systems. This would entail a nucleon-number dependent PC
which in turn, unfortunately, opens for the inclusion of four-body
forces in larger-mass nuclei and nuclear matter. Such nucleon-number
dependent PC schemes will be explored in our future
work.  \vspace{0.5cm}
\begin{acknowledgments}
  We thank U. van Kolck, B. Long, T. Papenbrock, J. Rotureau,
  M. S. Sanchez, G. Rupak and Y.-H. Song for useful discussions and
  suggestions. G.~H. acknowledges the hospitality of Chalmers
  University of Technology where most of this work was carried
  out. This work was supported by the European Research Council (ERC)
  under the European Unions Horizon 2020 research and innovation
  programme (Grant agreement No. 758027), the Swedish Research Council
  (Grant No. 2017-04234), the Office of Nuclear Physics,
  U.S. Department of Energy, under grants desc0018223 (NUCLEI SciDAC-4
  collaboration) and by the Field Work Proposal ERKBP72 at Oak Ridge
  National Laboratory (ORNL). The computations were enabled by
  resources provided by the Swedish National Infrastructure for
  Computing (SNIC) at Chalmers Centre for Computational Science and
  Engineering (C3SE), the National Supercomputer Centre (NSC)
  partially funded by the Swedish Research Council, the Innovative and
  Novel Computational Impact on Theory and Experiment (INCITE)
  program. This research used resources of the Oak Ridge Leadership
  Computing Facility located at ORNL, which is supported by the Office
  of Science of the Department of Energy under Contract
  No. DE-AC05-00OR22725.

\end{acknowledgments}




\appendix



\bibliography{eft_ref} 
\bibliographystyle{apsrev4-1}

\end{document}